\documentclass[a4paper,12pt]{article}
\pdfoutput=1
\usepackage{feynmp-auto}
\usepackage{amsmath, amsfonts, amssymb}
\usepackage{graphicx}
\usepackage{enumerate}
\usepackage{hyperref}
\usepackage{latexsym}
\usepackage{dsfont}
\usepackage{hepnicenames}
\usepackage{enumerate}
\usepackage{soul}
\usepackage[normalem]{ulem}
\usepackage{bbold}

\usepackage{units}
\usepackage{mathrsfs,graphicx,rotating,amsmath,amsfonts,mathtools,booktabs,amssymb,wasysym}
\usepackage{hyperref}\usepackage{slashed}
\usepackage[nosort]{cite}
\usepackage[table,xcdraw,dvipsnames]{xcolor}
\usepackage{bm}
\usepackage{graphicx}
\usepackage{multirow,multicol}
\hypersetup{colorlinks,bookmarksopen,bookmarksnumbered,
linkcolor=blus,pdfstartview=FitH,urlcolor=rossos,citecolor=verde}
\allowdisplaybreaks

\renewcommand{\[}{\left[}
\renewcommand{\]}{\right]}
\renewcommand{\(}{\left(}
\renewcommand{\)}{\right)}

\def\Lag{\mathscr{L}}

\newcommand{\mio}[1]{}

\def\bpm{\begin{pmatrix}}
\def\epm{\end{pmatrix}}

\usepackage{mathrsfs}

 \newcommand{\fig}[1]{~\ref{fig:#1}}
\newcommand{\sfrac}[2]{#1/#2}
 \newcommand{\One}{1\!\!\hbox{I}}

\newcommand{\Q}{{\cal Q}}

\allowdisplaybreaks
\usepackage{multicol}
\usepackage{color}
\definecolor{rosso}{cmyk}{0,1,1,0.4}
\definecolor{rossos}{cmyk}{0,1,1,0.55}
\definecolor{rossoc}{cmyk}{0,1,1,0.2}
\definecolor{blu}{cmyk}{1,1,0,0.3}
\definecolor{blus}{cmyk}{1,1,0,0.6}
\definecolor{bluc}{cmyk}{1,1,0,0.1}
\definecolor{verde}{cmyk}{0.92,0,0.59,0.25}
\definecolor{verdec}{cmyk}{0.92,0,0.59,0.15}
\definecolor{verdes}{cmyk}{0.92,0,0.59,0.4}

\newcommand{\riga}[1]{\noalign{\hbox{\parbox{\textwidth}{#1}}}\nonumber}

\newcommand{\eq}[1]{~{\rm (\ref{eq:#1})}}

\newcommand{\GeV}{\,{\rm GeV}}
\newcommand{\TeV}{\,{\rm TeV}}

\newcommand{\Tr}{\,{\rm Tr}}
\newcommand{\diag}{\,{\rm diag}}

\def\circa#1{\,\raise.3ex\hbox{$#1$\kern-.75em\lower1ex\hbox{$\sim$}}\,}

\newcommand{\beq}{\begin{equation}}
\newcommand{\eeq}{\end{equation}}

\newcommand{\bea}{\begin{eqnarray}}
\newcommand{\eea}{\end{eqnarray}}
\newcommand{\be}{\begin{equation}}
\newcommand{\ee}{\end{equation}}
\font\tenrsfs=rsfs10 at 12pt
\font\sevenrsfs=rsfs7
\font\fiversfs=rsfs5
\newfam\rsfsfam
\textfont\rsfsfam=\tenrsfs
\scriptfont\rsfsfam=\sevenrsfs
\scriptscriptfont\rsfsfam=\fiversfs

\newsavebox\MBox

\newcommand{\Sp}{\,{\rm Sp}}

\newcommand{\eV}{\,{\rm eV}}
\newcommand{\SU}{\,{\rm SU}}
\newcommand{\SO}{\,{\rm SO}}

\def\circa#1{\,\raise.3ex\hbox{$#1$\kern-.75em\lower1ex\hbox{$\sim$}}\,}
\makeatletter

\font\ital=cmu10

\def\hhref#1{\href{http://arxiv.org/abs/#1}{arXiv:#1}}
\usepackage{xstring}
\newcommand{\hhrefq}[1]{\IfSubStr{#1}{:}{\href{http://inspirehep.net/search?ln=en&ln=en&p=#1&of=hb&action_search=Search&sf=&so=d&rm=&rg=25&sc=0}{InSpire:#1}}{\hhref{#1}}}

\def\art{\@ifnextchar[{\eart}{\oart}}
\def\eart[#1]#2#3#4#5#6{{\rm #2}, {\em #3 \bf #4} {\rm (#6) #5} ({\em #1})}
\def\article{\@ifnextchar[{\earticle}{\oarticle}}
\def\oarticle#1#2#3#4#5#6{{\rm #1}, {\ital ``#6''}, {\rm #2 #3 (#5) #4}}
\def\earticle[#1]#2#3#4#5#6#7{{\rm #2}, {\ital ``#7''}, {\rm #3 #4 (#6) #5}  [\hhrefq{#1}]}
\def\hepart[#1]#2{{\rm #2, \sl#1}}
\def\heparticle[#1]#2#3{#2, {\ital ``#3''} [\hhrefq{#1}]}
\newcommand{\doi}[1]{\href{http://dx.doi.org/#1}{[link]}}

\newcommand{\hhrefqq}[1]{\IfBeginWith{#1}{10.}{\href{https://doi.org/#1}{doi:#1}}{\hhrefq{#1}}}
\def\earticle[#1]#2#3#4#5#6#7{{\rm #2}, {\ital ``#7''}, {\rm #3 #4 (#6) #5}  [\hhrefqq{#1}]}
\renewenvironment{thebibliography}[1]
     {\begin{multicols}{2}[\section*{\refname}]%
      \@mkboth{\MakeUppercase\refname}{\MakeUppercase\refname}%
      \list{\@biblabel{\@arabic\c@enumiv}}%
           {\settowidth\labelwidth{\@biblabel{#1}}%
            \leftmargin\labelwidth
            \advance\leftmargin\labelsep
            \@openbib@code
            \usecounter{enumiv}%
            \let\p@enumiv\@empty
            \renewcommand\theenumiv{\@arabic\c@enumiv}}%
      \sloppy
      \clubpenalty4000
      \@clubpenalty \clubpenalty
      \widowpenalty4000%
      \sfcode`\.\@m}
     {\def\@noitemerr
       {\@latex@warning{Empty `thebibliography' environment}}%
      \endlist\end{multicols}}

\newcounter{alphaequation}[equation]
\def\thealphaequation{\theequation\hbox to
0.6em{\hfil\alph{alphaequation}\hfil}}
\def\eqnsystem#1{
\def\@eqnnum{{\rm (\thealphaequation)}}
\def\@@eqncr{\let\@tempa\relax \ifcase\@eqcnt \def\@tempa{& & &} \or
  \def\@tempa{& &}\or \def\@tempa{&}\fi\@tempa
  \if@eqnsw\@eqnnum\refstepcounter{alphaequation}\fi
\global\@eqnswtrue\global\@eqcnt=0\cr}
\refstepcounter{equation} \let\@currentlabel\theequation \def\@tempb{#1}
\ifx\@tempb\empty\else\label{#1}\fi
\refstepcounter{alphaequation}
\let\@currentlabel\thealphaequation
\global\@eqnswtrue\global\@eqcnt=0 \tabskip\@centering\let\\=\@eqncr
$$\halign to \displaywidth\bgroup \@eqnsel\hskip\@centering
$\displaystyle\tabskip\z@{##}$&\global\@eqcnt\@ne
\hskip2\arraycolsep\hfil${##}$\hfil& \global\@eqcnt\tw@\hskip2\arraycolsep
$\displaystyle\tabskip\z@{##}$\hfil
\tabskip\@centering&\llap{##}\tabskip\z@\cr}
\def\endeqnsystem{\@@eqncr\egroup$$\global\@ignoretrue} \makeatother

\definecolor{Gray}{gray}{0.95}

\def\bal#1\eal{\begin{align}#1\end{align}}

\begin{document}
\vspace{1.5cm}

\begin{center}
{\Large\LARGE\Huge \bf \color{rossos}
% New gender of minima along cosets of composite scalars
%Coset cosmology of composite scalars
Coset Cosmology
%Cosmosettology
}\\[1cm]
{\bf Luca Di Luzio$^{a,b}$, Michele Redi$^c$, Alessandro Strumia$^{a}$, Daniele Teresi$^{a,b}$}\\[7mm]

{\it $^a$ Dipartimento di Fisica dell'Universit{\`a} di Pisa}\\[1mm]
{\it $^b$ INFN, Sezione di Pisa, Italy}\\[1mm]
{\it $^c$ INFN Sezione di Firenze, Via G. Sansone 1, I-50019 Sesto Fiorentino, Italy}\\[1mm]

\vspace{0.5cm}

{\large\bf\color{blus} Abstract}
\begin{quote}\large
We show that the potential of  Nambu-Goldstone bosons can have
two or more local minima e.g.\ at antipodal positions in the vacuum manifold.
This happens in many models of composite Higgs and of composite Dark Matter.
Trigonometric potentials lead to unusual features, such as symmetry non-restoration at high temperature.
In some models, such as the minimal $\rm SO(5)/SO(4)$ composite Higgs with fermions in the fundamental representation,  the two minima are degenerate giving cosmological domain-wall problems. Otherwise, an unusual cosmology arises, that can lead to supermassive primordial black holes;
to vacuum or thermal decays;
to a high-temperature phase of broken $\SU(2)_L$,  
possibly interesting for baryogenesis.
\end{quote}

\thispagestyle{empty}
\bigskip

\end{center}

%\DT{Un titolo pi\`u caDChy? Ad es.:``Composite (Higgs-)story'', ``The composite Universe'' (che per\`o assomiglia al paper di Pomarol), \ldots}

\setcounter{footnote}{0}

\tableofcontents

%\newpage

\section{Introduction}
Composite Higgs models (see e.g.~\cite{Dugan:1984hq,hep-ph/0412089,Agashe:2006at,hep-ph/0612048,1005.4269,1403.3116,1506.01961,1607.01659})
and composite models of Dark Matter (see e.g.~\cite{1005.0008,1204.2808,1503.08749,1602.07297,1703.06903,1704.07388,1707.05380,1801.06537,1812.04005}) received recent attention.
{In the introduction, for concreteness, we focus on composite Higgs models but our results apply in general to theories with spontaneous breaking of global symmetries.}
In order to partially justify the smallness of the electro-weak scale, Composite Higgs models assume that the Higgs doublet $H=(0,h)/\sqrt{2}$ is the pseudo-Nambu-Goldstone boson of some  approximate (possibly accidental) global symmetry $\mathscr{G}$ broken to a sub-group $\mathscr{H}$ at a scale $f$ by some strong dynamics, analogously to what happens with pions in QCD. The field space of composite scalars describes the bottom valley of the energy potential of the full theory,
well approximated by a coset with a non-trivial topology
e.g. a sphere. 
As a result, the low-energy effective field theory takes into account some effects beyond those of
low-energy renormalizable theories: the Higgs gauge and Yukawa interactions present in the Standard Model (SM) are corrected by trigonometric functions;
%with some model dependence and some more universal results dictated by group theory.
the potential, restricted for simplicity along the physical Higgs direction $h = (2 H^\dagger H)^{1/2}$
with period $2\pi f$ can be written as a Fourier series in $h^2$ 
\beq \label{eq:Vcos}
V(h) =\sum_{n=0}^\infty V_n \cos \frac{n h}{f}\eeq
with the higher-order terms being sub-leading.
If the lowest-periodicity term $V_1$ dominates, the potential has a single minimum:
this happens for pions in QCD. In composite Higgs models, instead, 
the Higgs vacuum expectation value (vev) $v$ must be somehow smaller than the compositeness scale $f$.
Then $V_1$ cannot dominate and higher order terms can generate extra local minima.
In many models $V_2$ dominates, giving rise to two nearly-antipodal local minima in the coset:
the SM minimum  at $h= v \ll f$, and an anti-SM minimum at $h\approx f\pi$.

%The non-dominance of $V_1$ is needed for phenomenological reasons:
%in view of experimental bounds the SM minimum must lie at $h = v \ll f$.
A compositeness scale $f$ larger than the Higgs vev $v$
comes at a price of a  tuning of order $f^2/v^2$. Therefore, there is a range of temperatures $ T \circa{<}  4\pi f$ where  quantum and thermal corrections
are computable in terms of low-energy degrees of freedom. This will allow us to compute the new interesting cosmological effects
related to the two minima in the Higgs potential.

An unusual feature specific of composite models  is that thermal corrections do not select $h=0$
`burning' the vacuum at $h \sim f\pi$: both minima remain present in the thermal potential. 
As we will see in the following, this leads to a number of interesting cosmological consequences.
Various groups studied cosmological implications of the Higgs as a pseudo-Goldstone boson \cite{hep-th/0107141,hep-ph/0409070,hep-ph/0607158,0706.3388,1007.1468,1104.4791,1110.2876,1508.05121,1803.08546,1804.07314},
focusing  on the electro-weak phase transition for applications to baryogenesis. Our work differs from these references 
as we do not modify the electro-weak phase transition around the tuned vacuum and we study the implications of minima 
existing at temperatures below the confinement phase transitions that gives rise to the Goldstone bosons.

The paper is structured as follows.
In section~\ref{models} we focus on the Higgs potential (including thermal corrections) in composite Higgs models.
In most models, the coset includes extra scalars.
In section~\ref{PG} we extend the discussion studying the potential in the full coset,
considering both composite Higgs models and composite Dark Matter models.
Cosmological implications are discussed in section~\ref{cosmo}.
Conclusions are given in section~\ref{concl}.

\section{Higgs potential in composite Higgs models}\label{models}

Composite Higgs models are often studied from a 
low-energy effective theory perspective,
as present experiments only offer bounds from this limited point of view.
An effective theorist can assume the
pattern of global symmetries $\mathscr{G}/ \mathscr{H}$ needed to get the desired 
phenomenological outcome.
Often, complicated constructions with extra custodial and other symmetries
are proposed in order to keep $f$ as low as possible,
as a $f \gg v $ comes with a fine-tuning of order $(f/v)^2$.
Given the bounds from LHC, we will not limit our study to a TeV-scale $f$, as a
much larger scale could arise for anthropic reasons.

\smallskip

The minimal model assumes an $\SO(5)/\SO(4)$ coset~\cite{hep-ph/0412089,hep-ph/0612048}:
no known confining gauge theory  in $3+1$ dimensions provides such symmetry structure.
One can wonder if the low-energy models might lie in a 4-dimensional swampland.
Constructions with warped extra dimensions reduce to effective theories with
$\SO(5)/\SO(4)$ structure, 
when described by an observer living on a 4-dimensional brane. 
As we will see, these models generate two vacua
at $h=0$ and $h=f\pi$. In appendix~\ref{ineq} we show that these
are distinct points in field space, like the North and South pole of the Earth.
The two vacua are non-degenerate in models with spinorial representations~\cite{hep-ph/0412089}, 
as they have double periodicity.  As such models are subject to strong 
constraints from electro-weak precision tests
of $b_L$ couplings,
ref.~\cite{Agashe:2006at,hep-ph/0612048} proposed models based on a 5 representation:
in these models minima are degenerate, giving rise to possible domain-wall problems in cosmology.
Lifting the degeneracy is difficult, because 
low-energy global symmetries are gauge symmetries in the warped extra dimensions.
%\DT{as well as \emph{any} $\mathscr{G}/ \mathscr{H}$ pattern with a partial gauging of a subgroup $G$.} \xxx{AS: non capisco}

\smallskip

A possible UV realisation  of composite Higgs models has been proposed in~\cite{1607.01659}.
This construction employs a new gauge interaction $G_{\rm DC}$ to generate the spontaneous breaking of global symmetries, and
elementary scalars to obtain the needed flavour structure through partially composite fermions.\footnote{The weak scale is as unnatural as in the SM, 
within the assumption that quadratic divergences indicate contributions of order of the Planck scale.
%Similar naturalness issues can arise in
%scenarios where flavour arises from 4-fermion operators:
%in order to reduce unseen new-physics flavour violations,
%theorists assume that operators involving two SM fermions and two new fermions acquire 
%lower nearly-renormalizable dimensions
%thanks to some unspecified new strong dynamics.
%The assumption that  two new fermions behave like a scalar
%brings both its good (for flavour) and bad (for naturalness) aspects.
}
This restricts the possible accidental global symmetries $\mathscr{G}/ \mathscr{H}$~\cite{1607.01659}:
for example $N_F$ `flavours' of techni-fermions
give $\SU(N_F)_L\otimes \SU(N_F)_R/ \SU(N_F)$ for $G_{\rm DC}=\SU(N_c)$ gauge groups;
$ \SU(N_F) / \SO(N_F)$ for $\SO(N_c)$ gauge groups;
$\SU(N_F)/ \Sp(N_F)$ for $\Sp(N_c)$ gauge groups.
Other composite particles do not lie in arbitrary representations.
Furthermore (as in QCD) the global symmetry can be broken by
dark-quark masses giving a specific UV-dominated contribution to the pseudo-Goldstone potential
that allows to remove the minimum at $h \approx f\pi$
or to make it non-degenerate.\footnote{Such terms might
vanish if one demands that all mass scales are dynamically generated.}

\medskip 

While models are sometimes complicated, their final results needed for our study can be 
understood in a simple way, as we now discuss.

\subsubsection*{Gauge interactions}
For  symmetric cosets the Nambu-Goldstone bosons (that include the Higgs doublet $H$)  can be parametrised with the unitary matrix
 $\mathscr{U} = \exp{(2i \Pi/f)}$ where $\Pi = T^{\hat a} \pi^{\hat a}$ are the broken generators. 
 %We normalize the generators such that $h=0$ is equivalent to $h=2\pi f$.
Their gauge-covariant kinetic term is\footnote{This 
applies to $\SU(N_F)_L\otimes \SU(N_F)_R/ \SU(N_F)$, $ \SU(N_F) / \SO(N_F)$, 
$\SU(N_F)/ \Sp(N_F)$.  For $\SO(N)/\SO(N-1)$ the pseudo-Goldstones can be parametrised by a vector $\Phi$ with fixed length and
kinetic term $f^2(D_\mu \Phi)\cdot (D^\mu \Phi)/2$, as e.g.~in section~\ref{SO6}.}
\beq \label{eq:gauge_cov} \frac{f^2}{4} \Tr[D_\mu \mathscr{U}^\dag \,D^\mu \mathscr{U} ]=
\frac{(\partial_\mu h)^2}{2}  + 
M_W^2(h) \[ W^+_\mu W^{-\mu} 
+ \frac{Z_\mu Z^\mu}{2 \cos^2\theta_{\rm W}}  \] +\cdots
\eeq
This Higgs boson $h$ is $2\pi f$ periodic in the coset but different periodicities for $M_W$ and, as we will see, for the potential are possible.
In the model of~\cite{Dugan:1984hq} based on $\mathscr{G}/\mathscr{H} = \mathrm{SU}(5)/\mathrm{SO}(5)$, one finds
\begin{eqnsystem}{sys:MWh}  \label{eq:Georgi}
\Pi &=& \frac{1}{2 \sqrt{2}} \begin{pmatrix}
0 & h & h & \cdots\cr
h & 0 &0& \cdots\cr
h & 0 &0& \cdots\cr
\vdots &\vdots &\vdots& \ddots
\end{pmatrix}+\cdots
\qquad\Rightarrow \qquad
M_W = g_2 f \sin \frac{h}{2f} \\
\riga{such that $M_W=0$ only at $h=0$.
In other models~\cite{hep-ph/0412089,hep-ph/0612048,1607.01659}}\\
 \label{eq:MMW1}
\Pi &=& \frac12 \begin{pmatrix}
0 & h & \cdots\cr
h & 0 & \cdots\cr
\vdots &\vdots & \ddots
\end{pmatrix}+\cdots
\qquad\Rightarrow \qquad
M_W = \frac{g_2 f}{2} \sin \frac{h}{f} 
\end{eqnsystem}
such that $M_W$ vanishes at $h=\{0, f\pi\}$. 
At the latter point the matrix $\mathscr{U}$ is diagonal with elements 
equal to either $1$ or $-1$, giving rise to the periodicity $\pi f$ in eq.~\eqref{eq:gauge_cov}. 
%Note that this is not the case instead for the matrix in \eqref{eq:Georgi}.
The above two functions $M_W(h)$ reduce to the SM expression for $h\to 0$ and have periodicities $2\pi f/N$
with $N=\{1,2\}$.
Other periodicities might arise in other models.

\subsubsection*{Yukawa interactions}
The top Yukawa interaction depends on extra group theory and model
details, such as the embedding of top quarks, and how many
insertions of $\mathscr{U} $ are necessary to obtain the top Yukawa interaction.
At the end, the various possibilities again simply correspond to the lowest coefficients in a Fourier series.
The SM expression of $M_t(h) t \bar t$ generalizes to
\beq \label{eq:Mtper}
M_t(h) = \left\{ \begin{array}{ll}
\displaystyle \frac{y_t f}{\sqrt{2}}  \sin\frac{h}{f}  & \hbox{in~\cite{1607.01659,hep-ph/0412089}} \cr 
\displaystyle\frac{y_t f}{2\sqrt{2}} \sin\frac{2h}{f}  & \hbox{in~\cite{hep-ph/0612048}} \cr
\displaystyle\frac{y_t f}{4\sqrt{2}} \sin\frac{4h}{f}  & \hbox{in~\cite{1808.10175}} 
\end{array}\right.
 \eeq
where, in each given model and coset, only one term is usually present.
Different periodicities might be possible in
fundamental composite Higgs theories, depending on the confining
gauge group~\cite{1607.01659}. 
The top mass vanishes at $h=0$ and, in the second {(third)} possibility, also at $h=f\pi$ {($f\pi/2$)},
with implications for quantum and thermal potentials.

\subsection{SM loop contributions to the Higgs potential}\label{VH}
The SM gauge couplings, $g_{2,Y}$, and the top Yukawa $y_t$  are sizeable and explicitly break the 
approximate global symmetry $\mathscr{G}$ generating at quantum level
the SM Higgs potential. The (often) 
dominant part of the Higgs potential can be roughly estimated, without doing any new computation, 
from the quadratically divergent part of the one-loop Coleman-Weinberg 
SM potential \cite{Coleman:1973jx}, replacing the SM expressions for $M_{W,Z,t}(h)$
with the coset-generalized masses $M_{W,Z,t}(h)$ given in the previous section,
and introducing two cut-offs $\Lambda_{\rm gauge}$ and $ \Lambda_{\rm top}$
of the order of the compositeness scale:
\begin{equation}\label{eq:VLambda2}
V(h) \approx \frac{1}{(4\pi)^2}\[
\frac{3}{2}(2 M_W^2(h) + M_Z^2(h))\Lambda_{\rm gauge}^2  -  6M_t^2(h) \Lambda_{\rm top}^2\]+\cdots \, .
\end{equation}
The $\cdots$ denote smaller low-energy terms of order $M_{W,Z,t}^4\ln M_{W,Z,t}^2$
as well as, crucially, extra breaking effects unrelated to the low-energy
SM couplings such as higher-order corrections to fermion kinetic terms.
These give significant contributions because $V$ is given by
power-divergent quantum corrections, as discussed in section~\ref{sec:VYukawa}.
By using formulas such as $\sin^2 x = (1-\cos2x)/2$ the potential is brought to the form of eq.\eq{Vcos}.
In the next sections we discuss the approximations that lead to eq.\eq{VLambda2}.

\medskip

The composite-Higgs thermal potential $V_T(h)$ can
be obtained at one-loop  from its SM expression (see e.g.~\cite{hep-ph/9901312}) with the same trick
of promoting the SM expressions for $M_{W,Z,t}(h)$ to their coset-generalized 
extensions:\footnote{For related studies of finite-temperature effects 
in the presence of pseudo Goldstone bosons see also~\cite{Gupta:1991ve,Holman:1992uj,hep-ph/0409070,1508.05121,1803.08546,1804.07314}.}
\be \label{eq:VT}
V_T(h) = \frac{T^4}{2 \pi^2} \[  6 J_B \(\frac{M^2_W }{ T^2}\) + 3 J_B \(\frac{M^2_Z }{ T^2}\) -12J_F \( \frac{M^2_t  }{ T^2}\) \].
\ee
The usual  bosonic and fermionic thermal
$J$ functions can be expanded in the high-$T$ limit as 
\begin{eqnsystem}{sys:J}
J_F (\epsilon) &= &\int_0^\infty x^2 \ln(1+e^{-\sqrt{x^2+\epsilon}}) dx  \simeq 
\frac{7 \pi^4}{360} - \frac{\pi^2}{24} \epsilon\\
J_B (\epsilon)  &=&
\int_0^\infty x^2 \ln(1-e^{-\sqrt{x^2+\epsilon}}) dx\simeq  - \frac{\pi^4}{45} + \frac{\pi^2}{12} \epsilon
\end{eqnsystem}
reducing to the usual thermal mass.
This is a good approximation around the minima: we see that the thermal corrections to the
potential give a minimum at all values of $h$ such that $M_{W,Z,t}=0$.
In many models this includes $h=f\pi$ together with $h=0$.

\subsubsection{Gauge contribution}\label{sec:gauge_contribution}
We now  discuss more precisely the gauge contribution to the potential. 
In the Landau gauge, only the transverse part of the effective gauge Lagrangian  contributes. 
The quadratic Lagrangian in momentum space (keeping only the transverse part) can be written as
\begin{equation}\label{eq:gauge}
\Lag_{\rm eff} \approx \frac{1}{2} \[ -p^2 + M_A^2(h) \] A_\mu  \(g^{\mu\nu} - \frac{p^\mu p^\nu }{ p^2}\) A_\nu\, ,
\end{equation}
where we neglected the momentum dependence of form factors originating from the 
strong dynamics\footnote{More precisely, we approximate the form factors with a 
step function $\Pi(E) \propto \theta(\Lambda - E)$.} 
and reabsorbed Higgs-independent terms into the renormalization of the gauge fields. 
The leading contributions to~\eqref{eq:gauge} originate from the gauge kinetic Lagrangian 
and from eq.~\eqref{eq:gauge_cov}. Higher-order corrections, including Higgs-dependent wave-functions, can be absorbed into the function $M_A(h)$ and, as such, are sub-leading with respect to the tree-level contribution in eq.~\eqref{eq:gauge_cov}. The Coleman-Weinberg potential obtained from eq.~\eqref{eq:gauge} is 
\begin{eqnsystem}{sys:VYuk}
\label{eq:Vgauge}
V_{\rm gauge} &\approx  &-\frac{i}{2}  \int \frac{d^4 p}{(2 \pi)^4} \[ 6 \ln\(p^2 - M_W^2(h)\) +  3 \ln\(p^2 - M_Z^2(h)\) \]
\label{eq:VgaugeT0}\\
&\approx& \frac{3}{2\, (4\pi)^2}
(2 M_W(h)^2 + M_Z(h)^2)\Lambda_{\rm gauge}^2  + \cdots
\end{eqnsystem}
which gives eq.~\eqref{eq:VLambda2}. The finite-temperature part of the potential can be obtained from the analogous of eq.~\eqref{eq:Vgauge}, in a well-defined and calculable way, since the momentum integrals are cut by $T \lesssim f \ll \Lambda$, obtaining eq.~\eqref{eq:VT}. 

\subsubsection{Yukawa contribution}\label{sec:VYukawa}
A more precise estimate of top-Yukawa power-divergent corrections can be obtained considering the 
general form of the effective Lagrangian for the top-quark sector and $h$.
Corrections $Z_{Q}$ and $Z_U$ to top quark kinetic terms can also be relevant 
(in the fundamental theory of~\cite{1607.01659}
this happens when their dark-Yukawa couplings are large enough), such that
\begin{align}\label{eq:Lagr}
\Lag_{\rm eff} &\approx
 \[1 + Z_Q (h)\] 
 \overline{Q} i \slashed{D}Q \,+\[1 + Z_U (h) \]
  \overline{U} i\slashed{D} 
 U  - \Big[ M_t(h) Q U + \hbox{h.c.} \Big] \;.
\end{align}
In the limit of interest $E \ll \Lambda$ we can neglect the momentum-dependence of the wavefunctions, as above.  
In this approximation we absorbed the Higgs-independent effects in the renormalization of the fermion fields. 
The functions $Z_Q(h), Z_U(h)$ are model-dependent trigonometric functions, 
multiplied by possibly small coefficients.
The expansion makes sense if they are sufficiently smaller than 1. 
At zero temperature, the Coleman-Weinberg potential obtained from~\eqref{eq:Lagr}
%\footnote{Up to ${\cal O}(1)$ corrections due to the neglecting of the momentum-dependence of the DC form factors or, more precisely, to their approximation as $const. \times \theta(\Lambda - p)$.} 
is
\begin{eqnsystem}{sys:VYuk}
\label{eq:VYukawa}
V_{\rm Yukawa} &\approx  &6 \, i \int \frac{d^4 p}{(2 \pi)^4} \left\{ \ln (1 + Z_Q) 
+ \;\ln\[p^2(1 + Z_Q )(1 + Z_U) - M_t^2\] \right\}
\label{eq:VYukawaT0}\\
&\approx& - \frac{12 \Lambda^4_{\rm top}}{(4 \pi)^2} \Big[2 \ln (1 + Z_Q ) + \ln (1 + Z_U )\Big] 
- \frac{6 \Lambda^2_{\rm top}}{(4 \pi)^2} \frac{ M_t^2 }{(1 + Z_Q )(1 + Z_U ) } + \cdots.
\label{eq:VYukawaT02}
\end{eqnsystem}
In the limit of negligible $Z_{Q,U}$ this reduces to eq.\eq{VLambda2}. 
In general, the term of order $\Lambda_{\rm top}^4$ dominates,
unless $Z_{Q,U} \circa{<} f^2/\Lambda_{\rm top}^2$.
In the models of~\cite{hep-ph/0412089,hep-ph/0612048} small $Z_{Q,U}$ are
required to obtain the phenomenologically interesting situation $v \ll f$. 
In some models in~\cite{1607.01659} $Z_{Q,U}$ can be neglected being
further suppressed by $m_{\mathcal{F}}/\Lambda$, $f^2/\Lambda^2$. 
%Therefore, according to the model, the potential of eq.~\eqref{eq:VYukawaT0} can generate one or some of the terms $A, B, C$ below.

\medskip

Again, the finite-temperature part of the potential can be obtained from the analogous of eq.~\eqref{eq:VYukawa} in a well-defined way. The terms depending on only the wavefunctions are proportional to $T^4$ and can be generically neglected for $T \ll f$, with respect to the term depending on $M_t^2$, that contains thermal-mass contributions $\mathcal{O}(T^2 f^2)$. We thus have:
%\begin{equation}\label{eq:VYukawaT}
%V_T|_{\rm Yukawa} = \frac{T^2 f^2}{8} \frac{ y_t^2 \phi(h/f)^2}{(1 + Z_Q )(1 + Z_U ) } \simeq \frac{T^2 f^2}{8} y_t^2 \phi(h/f)^2
%\end{equation}
\begin{equation}\label{eq:VYukawaT}
V_T|_{\rm Yukawa} = -\frac{6 \, T^4}{\pi^2} \, J_F \[ \frac{M_t^2}{(1 + Z_Q )(1 + Z_U )} \] \simeq -\frac{6 \, T^4}{\pi^2} \, J_F(M_t^2)
\end{equation}
that gives eq.~\eqref{eq:VT}. Therefore, for the thermal correction we only need to consider the different possibilities for the function $M_t(h)$.

\subsection{General parametrization of the Higgs potential}\label{Vparam}
We assume that the SM-like minimum lies at $h\ll f$.
Expanding eq.\eq{Vcos} in this limit (corresponding to $\mathscr{U}=\One$) the potential reduces to the SM form
\beq \label{eq:VSM}
V(h) \simeq  V(0) -\frac{M_h^2}{4}  h^2 + \frac{\lambda}{4} h^4+\cdots\eeq
with 
\beq \label{eq:SMpar}
V(0)=\sum_{n=0}^{\infty} V_n,\qquad
M_h^2 =\frac{2}{f^2}\sum_{n=1}^{\infty} n^2 V_n,\qquad
\lambda =\frac{1}{6f^4} \sum_{n=1}^{\infty} n^4 V_n.\eeq
Expanding around the antipode $h = \pi f$ 
%(corresponding ${\cal U}=-\One$) 
gives the same potential with $V_n\to (-1)^n V_n$.
Notice that $V(h)= V(h+2\pi f)=V(-h)$, so that $V(h)$ is fully characterised by its values in the
$0\le h\le f\pi$ domain.

Different models generate different combinations of the coefficients $V_{n}$.
The general structure of the potential can be found e.g.~in \cite{1105.5403}.
In general the functional form of the potential is fully determined by the couplings that explicitly break the global symmetry associated to
the Higgs boson. If generated to leading order in the strong sector coupling, the natural size of each contribution to the potential is
\begin{equation}
\frac{g_*^4 f^4}{16\pi^2} \epsilon^{i} \, ,
\end{equation}
where $g_*$ is the relevant strong sector coupling, $\epsilon$ parametrizes the breaking of the global symmetry and $i$ is the number of insertions required to
generate the contribution to the potential. For gauge couplings $\epsilon=g_{\rm SM}/g_*$ and $i=2$, 
while for Yukawa couplings it is model dependent. In particular, for models with partially 
composite fermions there are more couplings than in the SM that break the global symmetry and consequently the contributions to the potential cannot be related to SM Yukawa couplings 
in general. For our purposes it is sufficient to include the Fourier terms with $n=\{1,2,4\}$ in the Higgs potential of eq.\eq{Vcos}.\footnote{Matching to more standard notations 
(as e.g.~in \cite{1005.4269}), the potential including the lowest Fourier modes 
can be also written as 
$V(h) = \alpha \cos \sfrac{h}{f} - \beta \sin^2 \sfrac{h}{f} + \gamma \sin^4 \sfrac{h}{f}$,  
with $\alpha=V_1$, $\beta=2(V_2+4V_4)$, $\gamma=8V_4$.
}  
Focusing on $h$, a generic kinetic term can be made canonical 
through a field redefinition that affects $V_4$ and higher-order terms in the potential.

\begin{figure}[t]
$$\includegraphics[width=0.68 \textwidth]{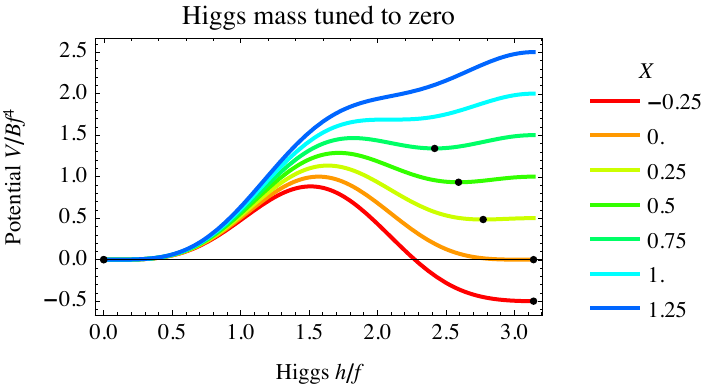}$$
\caption{\em\label{fig:V0} Possible composite Higgs potentials with lowest-frequency terms
for different values of the free parameter $X$ defined in eq.\eq{R}.}
\end{figure}

The lowest frequency  $V_1$ is generated with large coefficient $V_1/f^4 \sim g_2^2 g_*^2/(4\pi)^2$
by SM gauge interactions in the model of~\cite{Dugan:1984hq} (see eq.\eq{Georgi} and eq.\eq{VLambda2}).
In other models it can be generated with small coefficients: in the fundamental theories of~\cite{1607.01659} it arises
proportionally to dark-fermion masses as $V_1/f^4 \sim M_{\cal Q} \Lambda/f^2$.
In models with partial compositeness $V_1$ is different from zero only in the presence of specific representations~\cite{1210.7114}, e.g.\ a $4$ if $\mathscr{G}/\mathscr{H} =\SO(5)/\SO(4)$~\cite{hep-ph/0412089}.

In the models of~\cite{hep-ph/0412089,hep-ph/0612048,1607.01659} 
where $M_W^2$ is given by eq.\eq{MMW1}, SM gauge interactions  generate $V_2/f^4 \sim g_2^2/(4\pi)^2 g_*^2$ and subleading terms $V_4/f^4 \sim  g_2^4/(4\pi)^2$. 
The top Yukawa coupling contributes in similar ways, depending on its periodicity in eq.\eq{Mtper} and typically dominates numerically.
%The condition $V_2=0$  needs $8y_t^2\Lambda_{\rm top}^2=(3g_2^2+g_Y^2)\Lambda_{\rm gauge}^2$. 
%Then writing the coefficients in the form
%\beq  V = V_0 - Cf^4 \cos\frac{h}{f}-\frac{A+B}{2}f^4\cos \frac{2h}{f} + \frac{B}{8} f^4\cos \frac{4h}{f}\eeq
%and tuning $V_0$ such that $V(0)=0$ gives the convenient expression
%\beq\label{eq:VABC}
%%\Lag = \frac{(\partial_\mu h)^2}{2} -V(h)\qquad
%V(h) = f^4 \bigg( A \sin^2\frac{h}{f} + B \sin^4\frac{h}{f} +C(1- \cos\frac{h}{f}) \bigg)\eeq
%In the models of~\cite{hep-ph/0412089,hep-ph/0612048,1607.01659}  the gauge sector contributes
%as $A \sim g_2^2/N$ and $B \sim g^4/(4\pi)^2$. 
%The top Yukawa coupling contributes as $A \sim y_t^2$ and $B \sim y_t^4/(4\pi)^2$ in the models of~\cite{1607.01659,hep-ph/0412089}.
%%\DT{Add discussion about naive expectations, tuning and too heavy Higgs; therefore often 3 frequencies.} 

Phenomenological considerations impose
that:
\begin{itemize} 
\item $v^2 \ll f^2$ i.e.~the tuning 
$V_1 +4 V_2 + 16 V_4 \ll \lambda f^4$. 

\item the SM Higgs quartic equals $\lambda \approx 0.086$ when renormalized at $2\TeV$ in the $\overline{\rm MS}$ scheme~\cite{1307.3536}.
This puts some pressure on composite Higgs models that often favour larger values. 
Already the gauge contribution gives a too large quartic unless $g_*\lesssim 4$ while the top Yukawa indicates $g_*\sim 2$.
In order to reproduce the Higgs mass either $g_*$ is small or some deviation from naive scaling must be assumed.
\end{itemize} 

The form of the tuned potential (up to an overall rescaling, if the value of $\lambda$ is ignored)
is determined by the free parameter 
\beq X = -V_1/8V_4.\label{eq:R}\eeq
%%We see that  $h = 0$ is a local minimum for $A<C/2$ and a maximum otherwise.
%$to -C$. So $h \simeq \pi f$ is a local minimum for $A>C/2$ and a maximum otherwise.
%
%As $M_h^2 \ll f^2$ is needed for phenomenological reasons, we tune $2A+C=0$.
%Up to an overall rescaling of the potential, its form is then fully determined by $C/B=-V_1/8V_4$:
Fig.~\ref{fig:V0} shows the various possibilities.
The SM minimum is the global minimum for $X>0$ and is a local minimum
for $X<0$.
For $X=0$ it is degenerate with the extra minimum at $h=f\pi$.
For growing $X$ such extra minimum shifts towards smaller $f$ and
finally disappears for $X>1$. 
Minima remain degenerate when the Higgs mass is not tuned to zero, 
since for $X=0$ the potential is symmetric around $h = f \pi/2$.
%Notice that the value of $B$ is determined by the measured value of the Higgs quartic $\lambda$.

In conclusion, an interesting structure of non-degenerate minima arises for $V_1 \lesssim  V_2\sim V_4$. 

\section{Potential of multiple pseudo-Goldstone bosons}\label{PG}
The previous discussion considered the potential along the Higgs direction, 
with extra pseudo-Goldstone bosons set to zero.
If present, they can qualitatively
change the conclusions: connecting the Higgs minima through different trajectories;
give new minima, etc. Singlets neutral under the SM gauge interactions can be especially light and relevant\footnote{A light dilaton could play a similar role as discussed in \cite{1803.08546,1804.07314}, however we focus on particles belonging to the coset.}. 
Their presence and potential is model dependent.
In section~\ref{SO6} we consider the  next-to-minimal composite Higgs model.
In section~\ref{QCD} we consider fundamental models based on $\SU(N_c)$ strong gauge interactions.

\subsection{Composite Higgs with SO(6)/SO(5)}\label{SO6}
The next-to-minimal composite Higgs model is based on the symmetry breaking pattern 
$\mathscr{G}/\mathscr{H}=\SO(6)/\SO(5)$~\cite{Gripaios:2009pe,Redi:2012ha}. 
The 5 Goldstone bosons $H$ and $\eta$ 
can be described by a real vector $\Phi$ with 6 components and fixed length $f$.
The electro-weak symmetry group acts on its first four components.
In the unitary gauge $H$ reduces to $h$,
and the coset is conveniently parametrised in terms of two spherical angles $\varphi$ and $\psi$ that depend on
$h$ and $\eta$
\begin{equation}
\Phi= f \left(0,~0,~0,~\sin {\varphi} \cos  {\psi},~\sin {\varphi} \sin  {\psi},~\cos {\varphi} \right)\,
\end{equation}
so that 
\begin{equation}
M_W= \frac {g_2 f} 2 \sin {\varphi} \cos {\psi} \,.
\end{equation}
Fermion masses are model dependent. For composite fermions in the 6 of SO(6) a unique embedding of $t_L$ exists while $t_R$
can couple to two different singlets corresponding to the fifth and sixth components of a vector. Denoting with $\alpha$ the angle one finds
\begin{equation}
M_t = \frac {y_t f} {\sqrt{2}} \sin {\varphi} \cos {\psi} \big[i \cos\alpha \sin {\varphi} \sin{\psi} +  \sin\alpha\cos {\varphi}\big] \,.
\end{equation}
The potential generated by SM gauge interactions  has the form
\begin{equation}\label{eq:potSO6}
\begin{split}
V(\varphi,\psi) \approx &~c_1 \sin^2 {\varphi}  \cos^2 {\psi}  + c_2 \sin^2 {\varphi} \big(\sin^2\alpha-\cos^2\alpha \sin^2 {\psi} \big)+\\
 &-c_3 \sin^2 {\varphi} \cos^2 {\psi} \big[\cos^2\alpha \sin^2 {\varphi} \sin^2{\psi} + \sin^2\alpha\cos^2 {\varphi} \big]
\end{split}
\end{equation}
where $c_{1}$ is generated by gauge and top  left couplings; 
$c_{2}$ by top right couplings; $c_3$ by the top Yukawa. Thus, the Yukawa contributions correspond respectively to the first, second and third terms in eq.~\eqref{eq:VYukawaT02}.

\begin{figure}[t]
$$\includegraphics[scale=1]{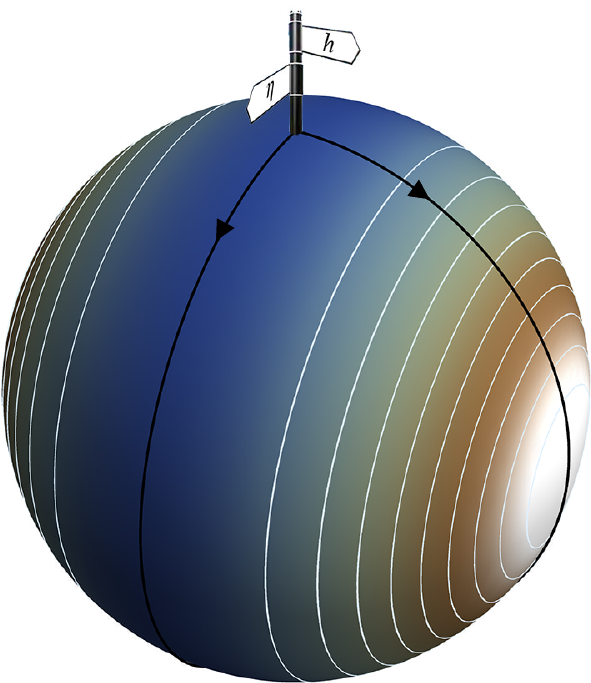} \qquad \qquad \qquad
\includegraphics[scale=1]{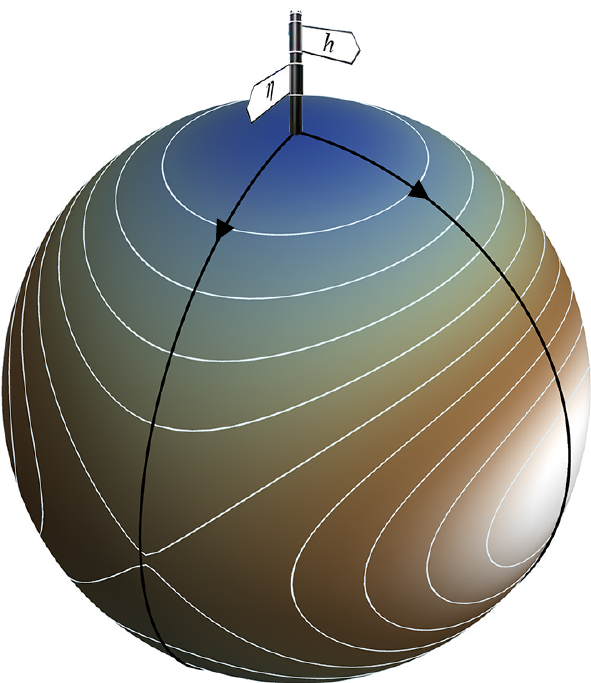}$$
\caption{\em The coset of the next-to-minimal composite Higgs model forms a sphere parameterized by
the Higgs $h$ and $\eta$ scalars.
The $h$ and $\eta$ directions are indicated on the North pole, which corresponds to $\Phi = \diag(0,0,0,0,0,f)$.
Contour lines of the potential $V(h,\eta)$ in eq.~\eqref{eq:potSO6} for $c_i=1$ and
for $t_R$ couplings $\alpha=\pi/4$ ($\alpha=\pi/2$) in the left (right) panel. The potential increases going from blue to brown to white.\label{fig:Palla} }
\end{figure}

Along $\psi=0$ we have $\varphi = h/f$ and 
the potential is identical to the potential of the minimal composite Higgs, with its two anti-podal minima
at $\varphi=0,\pi $. 
Increasing $\psi$ the potential barrier gets parametrically smaller, by an  amount that depends
on the model-dependent parameter $\alpha$. For $\alpha=\pi/4$ the barrier disappears along the direction $\psi=\pi/2$.
In this limit the singlet is an exact Goldstone boson and the antipodal points are connected through a valley of minima, as shown in the left panel of Fig.~\ref{fig:Palla}.

The singlet $\eta$ is anomalous under QCD behaving as an electro-weak axion (unless $f \gg v$). 
Therefore a breaking of its shift symmetry is  phenomenologically necessary; a barrier between the two minima is present for $\alpha \neq \pi/4$, as shown in the right panel of fig.~\ref{fig:Palla}.
%The previous analysis is however modified by the existence of the singlet. This can be seen most clearly  
%in the limit where $\alpha=\pi/4$ where the singlet becomes an exact GB. The potential exactly vanishes 
%for $\psi=\pi/2$ showing that the antipodal points are connected by a valley of minima. In this case the vacuum manifold 
%is a circle. 
%The domain walls discussed in SO(5)/SO(4) is therefore unstable and instead stable strings exist. \xxx{Discussion on strings?}
%For $\alpha\approx \pi/4$ the barrier along the singlet direction is smaller that the Higgs barrier. 
The two minima are degenerate: a small splitting can be obtained 
breaking the $\mathbb{Z}_2$ symmetry, for example coupling the SM fermions to a 4 of SO(6).
%becomes a 2 field problem. We do not expect significant differences with respect to single field analysis in this case.

This example shows a general phenomenon: in the presence of extra pseudo-Goldstone bosons
the topology of the vacuum can change, connecting minima along new paths. Bosons charged under $G_{\rm  SM}$ typically 
acquire potential barriers due to gauge loops, 
so they are not expected to change the qualitative features of the Higgs barriers 
but possibly introducing new local minima. 
Singlets on the other hand can have a small potential since 
their couplings to SM fermions are model dependent:
if they (approximately) preserve $\mathscr{G}$,
the local minima connected by $\eta$ can dominantly tunnel along the singlet direction rather than through the Higgs barrier. The opposite is obtained if the barrier along the singlet direction is large enough. The intermediate situation with comparable barriers requires a multi-field treatment.

\subsection{QCD-like theories}\label{QCD}
In order to extend the discussion to fundamental theories with multiple Goldstone bosons,
we focus on those based on a `dark-color' strong $\SU(N_c)$ gauge group with $N_F$ `dark-flavours'
of dark-quarks in the (anti)fundamental  of  $\SU(N_c)$, collectively denoted as $\Q$.
We assume that dark-quarks are charged under the SM gauge group $G_{\rm SM}$ forming
a vector-like representation  such that they can have masses $M_\Q$
and the new strong dynamics does not break $G_{\rm SM}$.
These  theories have been studied to construct models where dark matter is an accidentally
stable bound state of the new strong dynamics~\cite{1503.08749}. 
Furthermore, theories of composite Higgs are obtained in the presence of extra `dark scalars' ${\cal S}$~\cite{1607.01659}: we here assume that these do not lead to extra Goldstone bosons.

The  coset $\mathscr{G}/\mathscr{H}=\SU(N_F)_L\otimes\SU(N_F)_R/\SU(N_F)_V$ can be parametrised by a unitary matrix $\mathscr{U}$ with unit determinant and
thereby has the same topology as $\SU(N_F)$.
Up to ${\cal O}(E/f)^2$  the low energy effective Lagrangian is
\begin{eqnsystem}{sys:SU}\label{lagrangianN}
\mathscr{L}_{\rm eff} &=&\frac{f^2}{4}{\rm Tr} [D_\mu \mathscr{U} \, D^\mu \mathscr{U}^\dagger]-(  V_{\rm mass}+
V_{\rm gauge}+V_{\rm Yukawa})\\
V_{\rm mass} &=& -g_* f^3 {\rm Tr}[e^{i {\theta}/N_F} M_\Q \mathscr{U}^\dagger+\hbox{h.c.}] \\
V_{\rm gauge} &\approx& -\frac {3  g_*^2 f^4 } {2(4\pi)^2} \sum_{b} g_b^2 {\rm Tr}[\mathscr{U} T^b \mathscr{U}^\dagger T^b]=\frac {3  g_*^2 f^2 } {(4\pi)^2}  \Tr M_V^2+\hbox{cte} \,.
\end{eqnsystem}
$V_{\rm mass}$ is generated by constituent masses $M_\Q =\text{diag}(M_{\Q_1}, \ldots, M_{\Q_{N_F}})$, and
$V_{\rm Yukawa}$ by Yukawa interactions in the fundamental theory,
either with the SM Higgs or with dark-colored scalars ${\cal S}$  (as needed to get SM fermion masses in
theories of composite Higgs~\cite{1607.01659}). The $\SU(N_c)$ gauge theory can have a non-vanishing $\theta$ angle. 
Its effects can be included rotating $\theta$ to the dark quark mass matrix, that becomes
$\tilde M_\Q= e^{i {\theta}/N_F} M_\Q$ with $M_\Q$  a diagonal matrix with positive entries.
$V_{\rm gauge}$ is proportional to the squared  mass matrix $M_V^2$ of gauge bosons generated by the $\mathscr{U}$ background. 
The generators of the SM gauge group are $N_F\times N_F$  matrices $T^b$ 
determined by the SM gauge quantum numbers of $\Q$;
$g_b$ are the SM gauge couplings, and $g_*\sim 4\pi/\sqrt{N_c}$ is the effective strong coupling.

\smallskip

To study the vacuum of the theory we consider first the gauge contribution, as it
satisfies general properties:
$V_{\rm gauge}$ is minimal for configurations that do not break the gauge group $G_{\rm SM}$~\cite{Witten:1983ut}.
This implies that the minima of the gauge potential correspond to unitary matrices $\mathscr{U}$ 
block diagonal over each $G_{\rm SM}$ representation in $\Q$.
The $N_F$ centers of the flavour group
\beq \mathscr{U}_n=e^{2\pi i n /N_F}\diag(1,\ldots,1)\qquad\hbox{for integer}\qquad n=\{0,\ldots, N_F-1\}\eeq 
are always minima of $V_{\rm gauge}$. 
%These  minima of $V_{\rm gauge}$ can be reached moving in dark-pion $\pi$ field space from the origin $\mathscr{U}_0$
%along diagonal generators with non-vanishing entries.
%For example, for $N_F=2$ the generator $\diag(1,-1)$ connects all centers.
%For $N_F=3$ the generator $\diag(2,-1,-1)$ connects all centers.
%For $N_F=4$ the generator $\diag(3,-1,-1,-1)$ connects all centers, while $\diag(1,1,-1,-1)$ misses some of them.
%For $N_F=5$ the generators $\diag(4,-1,-1,-1,-1)$ and $\diag(2,2,2,-3,-3)$  connect all centers.
%For $N_F=6$ only $\diag(5,-1,-1,-1,-1,-1)$ connects all centers.
In addition if $\Q$ consists of $r$ reducible representations under $G_{\rm SM}$, the dark pions
include $r-1$ singlets, named $\eta$'s  (the $r$-th singlet being the heavy $\eta'$),  whose generators are block diagonal traceless matrices, and that do not receive mass from $V_{\rm gauge}$. 
Extra singlets  exist if  $\Q$ includes multiple copies of the same
representation.

\begin{figure}[t]
$$\includegraphics[width=0.5 \textwidth]{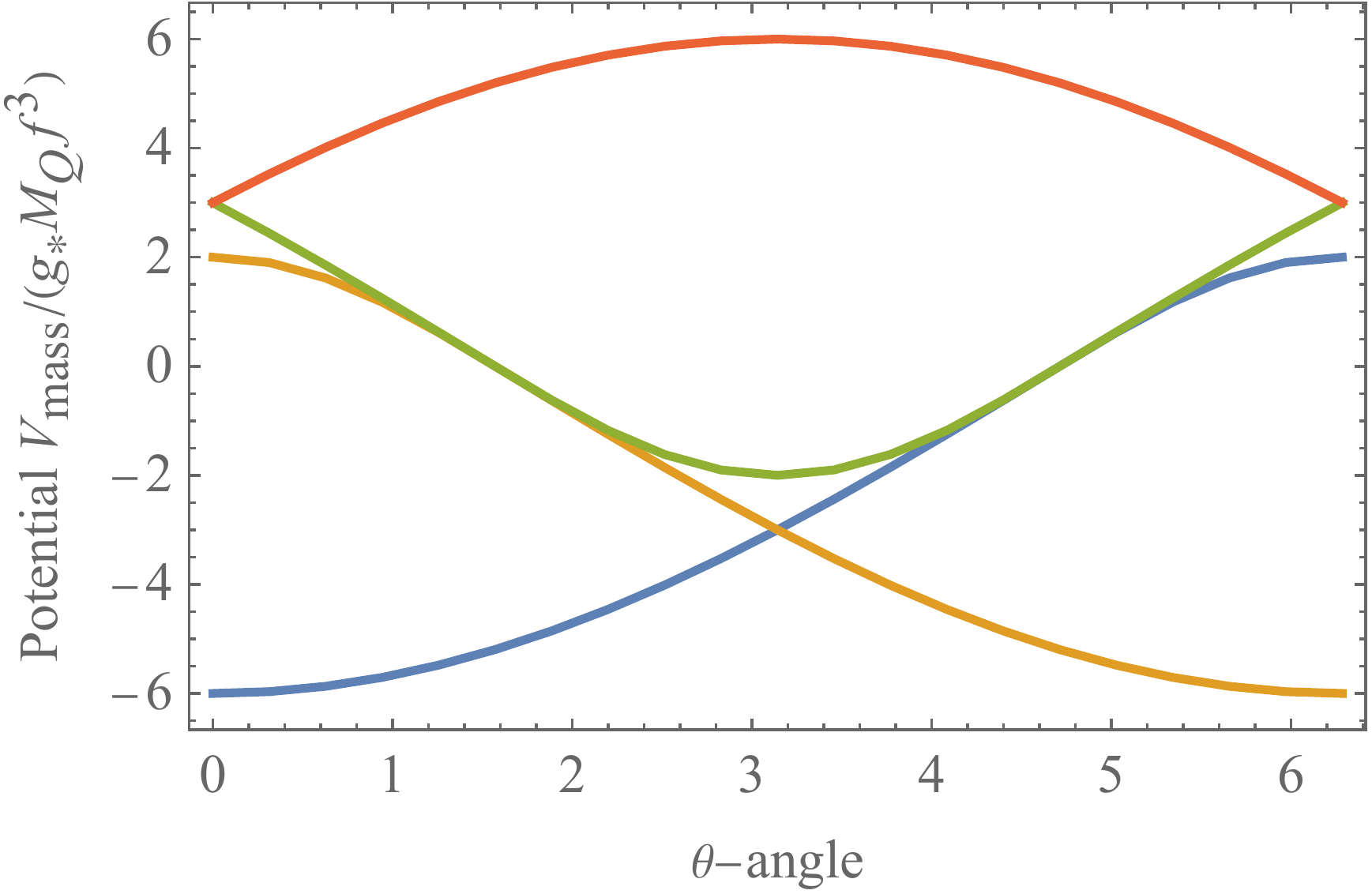}$$
\caption{\em\label{fig:Vtheta} Energy of the stationary points of the pseudo-Golstone
potential generate by dark quark masses,
eq.~\eqref{eq:VmassTheta}, in a theory with 3 degenerate flavours.
The two minima along $\eta$ (blue and yellow lines) level-cross at $\theta=\pi$
so that the ground state  has a cusp singularity. For $ \pi/2<\theta < 3/2 \pi$ they are minima of the full potential.  Green and red lines 
are always maxima.}
\end{figure}

\smallskip

Mass terms and Yukawa couplings can lift the degeneracy of singlets leading to local minima separated by potential barriers.
In appendix~\ref{QCDapp} we explicitly compute models with $N_F=2$ and $3$ flavours, finding a variety
of behaviours: there is only one minimum
in some models for some values of their parameters (this is the case of QCD);
in some cases there are valleys of minima, in some other cases there are multiple local minima separated by potential barriers.

For what concerns the minima generated by mass terms the discussion depends crucially on the $\theta$ angle in the dark sector, see appendix~\ref{theta2} for a derivation and more details. 
Let us consider, for simplicity, a theory with 3 flavours and degenerate masses $M_\Q$, and focus
on the pseudo-Goldstone boson $\eta$ associated to $\lambda^8$ (normalised as
$\mathscr{U}=\exp(i \eta \lambda^8/\sqrt{3}f)$ such that its periodicity is $2\pi f$).
Keeping all other dark pions at the origin, its potential is
\be \label{eq:VmassTheta}
 V_{\rm mass}(\eta)_\theta = -2 g_* f^3 M_{\Q} \[ 2 \cos \(\frac{\eta}{f}+\frac{\theta}{3}\) + \cos \(\frac{2 \eta}{f} - \frac{\theta}{3}\) \].
\ee
Although not manifest at first sight, physics is periodic in $\theta$ with period $2 \pi$ since
\be \label{eq:monodromy}
V_{\rm mass}( \eta)_\theta  = V_{\rm mass} \left( \eta - \frac{2\pi}{3}f\right)_{\theta + 2 \pi} \;.
\ee
Therefore a rotation $\theta \to \theta + 2 \pi$ corresponds to a shift in the compact field $\eta$. The potential~\eqref{eq:VmassTheta} for $\theta=0$ has a global minimum at $\eta=0$, a local minimum at $\eta = \pi f$ and two degenerate maxima at the other two centers $\eta/f = 2 \pi/3, 4 \pi/3$. Since a shift $\theta \to \theta + 3 \pi$ corresponds to flipping the sign of the constituent masses $M_\Q \to - M_{\Q}$ (and therefore $V_{\rm mass} \to - V_{\rm mass}$), for $\theta = \pi$ the potential~\eqref{eq:VmassTheta} has two degenerate minima. These are shifted to the centers  $\eta /f = 0, 4 \pi/3$ because of eq.~\eqref{eq:monodromy}, whereas there is a maximum at $\eta /f =2 \pi/3$. For  $\theta\approx \pi$ the two minima get split by $\Delta V\sim g_* f^3 M_{\Q} (\theta-\pi)^2$.

The potential of the stationary points as a function of $\theta$ is shown in fig.~\ref{fig:Vtheta}. Starting from $\theta=0$ the energy of the two minima along $\eta$ (blue and yellow lines) gets closer to each other until they cross at $\theta=\pi$, with the two points remaining distinct. 
On the other hand, the local minimum (yellow line) and one of the maxima (green line) merge into a single point at $\theta=\pi/2$, which is a saddle point along $\eta$. For $\pi/2 < \theta < 3\pi/2$ the minima along $\eta$ are true minima of the full $V_{\rm mass}$.

Summarising, for negligible gauge and Yukawa contributions $V_{\rm mass}$ has two degenerate minima for $\theta=\pi$, which become non-degenerate for $\theta\neq \pi$. 
In the presence of gauge or Yukawa interactions, multiple minima (degenerate or not) can exist for any value of $\theta$.

\subsection{Theories with the Higgs and composite scalars}
\label{partiallycomposite}

Finally, it is interesting to study what happens in theories that feature the SM elementary Higgs doublet $H$ 
together with pseudo-Goldstone bosons. We will see that the Higgs can participate in their possible multiple minima,
giving rise to multiple vacua that break differently the electro-weak group.

In models where the new constituent fermions $\Q$ have Yukawa couplings $\Q \Q H$, 
the pseudo-Goldstone bosons include a scalar $\pi_2$ with the same
quantum numbers as the Higgs.
The light Higgs doublet is in general a linear combination of $H$ and $\pi_2$  \cite{Antipin:2015jia,Agugliaro:2016clv,1609.05883}. 
If the mixing is large, it has a phenomenology similar to a composite Higgs.

% models where the vacuum misalignment is produced by the elementary 
%Higgs. If the strong sector has multiple minima the tuning of the electro-weak VEV can be done around one the minima while the other 
%will remain untuned. Thus this leads to a theory with different Higgs VEVs.

For example, we consider a model with $N_F=3$ where $\Q=\Q_L\oplus \Q_N$ has the same $G_{\rm SM}$
quantum numbers of the SM lepton doublet $L$ and of a right-handed neutrino $N$~\cite{Antipin:2015jia,Barducci:2018yer}. 
The fundamental Lagrangian contains
\begin{equation}
\mathscr{L}= y H \Q_N \Q_L^c+ \tilde{y} H^\dagger \Q_N^c \Q_L  + M_{\Q_N} \Q_N \Q_N^c + M_{\Q_L} \Q_L \Q_L^c + \hbox{h.c.}+\cdots
\end{equation}
Expanding the low-energy effective Lagrangian around the origin, $H$ mixes with $\pi_2$ 
\begin{equation}
\mathscr{L}_{\rm eff}=  -M_{\pi_2}^2 |\pi_2|^2 - i \sqrt{2} (y-{\tilde y}^*) g_* f^2  (  \pi_2^\dagger H+ \hbox{h.c.}) +\cdots
%+  y_+ g_* f \left(a_1 \eta \pi_2^\dagger H + a_3 \pi^a \pi_2^\dagger \sigma^a H + h. c.\right)
\label{mixing}
\end{equation}
The mixing parameter
\begin{equation}
\epsilon\equiv i \sqrt{2} \, (y-{\tilde y}^*) \frac{g_* f^2}{M_{\pi_2}^2},\qquad
M^2_{\pi_2}\approx    \frac{9 g_2^2 g_*^2}{4(4\pi)^2}f^2+ 2(M_{\Q_L} \cos \phi_L + M_{\Q_N} \cos \phi_N) g_* f
\label{eq:mixing}
\end{equation}
controls the degree of compositeness of the light Higgs.
For $\epsilon \ll 1$ the  light Higgs is mostly elementary, and the mixing contributes to its mass,
$\Delta V= - M_{\pi_2}^2 |\epsilon|^2 |H|^2$.
As a consequence $v \ll f$ can only be tuned around a single minimum of the strong  sector; at the other minima the weak gauge symmetry can be preserved
or badly broken.
%integrating out the composite Higgs. This produces a negative effective mass for the elementary Higgs,
%\begin{equation}
%\delta V= - M_\pi_2^2 |\epsilon|^2 |H|^2
%\end{equation}
%This negative shift of the Higgs mass parameter should be tuned with the bare $H^2$ contribution in order to reproduce the 
%electro-weak VEV. Around a different vacuum of the strong dynamics the Higgs VEV will not be tuned and the electro-weak symmetry might be 
%preserved or broken at a high scale. \xxx{Is the symmetry broken or unbroken in the falso vacuum?}
For $\epsilon \gg 1$ the light Higgs is mostly composite and the electro-weak symmetry is broken if the
mass matrix has a negative eigenvalue. Also in this case the tuning can be enforced around a minimum,
while the other induces a second local minimum of the composite Higgs.

% and the model above can be regarded as a UV completion of the composite Higgs. 
%The electro-weak symmetry in this case can be broken even though the diagonal mass entries are positive through the large mixing terms that give,
%\begin{equation}
%{\rm Det}[M^2]= M_H^2 M_\pi_2^2 - M_\pi_2^4 |\epsilon|^2
%\end{equation}
%Tuning of the electro-weak VEV can be realized imposing ${\rm Det}[M^2]\approx 0$ and negative around one of the minima of the strong sector. 
%Around a different minimum the tuning will be ineffective realizing effectively multiple Higgs minima.

\section{Cosmology}\label{cosmo}
In the previous sections we found that, in theories where the scalar field space has a non-trivial topology,
the potential can have multiple local minima which can be degenerate or not; separated by potential barriers or not.
We here explore the consequent cosmology.

A phase transition is expected to happen  at a critical temperature $T_{\rm cr} \sim \Lambda \sim 4\pi f$,
below which the Higgs exists as a composite particle. This physics depends on the strong dynamics, which is model dependent.
In QCD-like models the transition is expected to be weakly first order or cross-over. 
A special class of models
%, dual to walking technicolor, 
features a light dilaton and can be controlled
through holography that indicates a strong first order phase transition \cite{hep-th/0107141, 0706.3388,1104.4791,1711.11554}.
Due to the light dilaton, in some models there is no sharp distinction 
between the electro-weak  and  confinement phase transitions~\cite{1803.08546,1804.07314}.

We focus on lower temperatures, where the possible presence of two or more
inequivalent local minima in the Higgs potential
can leave cosmological signatures. 
The Higgs potential at finite temperature is reliably computable in the effective theory roughly up to temperatures $T \circa{<}\Lambda \sim 4\pi f$.\footnote{It is possible that the critical temperature is numerically around $f$, this is indeed what happens in QCD. In such a case, the thermal potential is calculable for $T \circa{<} f$. 
This is sufficient to study the fate of the minima of the potential at finite temperature.}
Thermal corrections to the potential due to a particle $X$ are of order $V_T \lesssim T^2 M_X^2(h)$, which has to compete with the zero-temperature potential $V \sim (\Lambda/4 \pi)^2 M_X^2(h)$;
therefore, they are significant at $T \circa{>} f$ in the whole coset,
and at $T\circa{>} M_h$ locally around the SM minimum where the curvature is tuned to be small.
%are significant at $T \circa{>} f$ in the whole coset,
%and at $T\circa{>} M_h$ locally around the SM minimum with tuned $M_h \ll f$.

Usually, thermal corrections to the SM Higgs potential
are roughly approximated by a thermal mass ${\cal O}(T)^2 h^2$
that selects $h=0$ as the only minimum, 
as this is the vacuum expectation value that makes massless 
the $W,Z,t$ particles that interact with the Higgs boson.

%thermal corrections select $h=0$ as the only minimum of
%the finite-temperature potential, because 
%SM vectors and fermions are massless at $h=0$.

The case of a pseudo-Goldstone boson is special: in the symmetric limit all points in the coset are equivalent,
and interactions that break the accidental global symmetry have a more complex structure
that allows for extra local minima.
The part of the potential generated by interactions with heavy particles receives negligible thermal corrections.
The part of the potential induced by interactions with light SM particles receives special thermal corrections.
Focusing, for simplicity, on the Higgs direction, multiple minima can arise around those field values 
at which the $W,Z$ and/or the top quark become massless.
In such a case the thermal potential given in eq.\eq{VT} 
can have the same multiple minima, separated by increasing barriers at large $T$.

%
%In our case SM vectors and fermions are massless at 
%such that multiple local minima can also be present in the thermal potential.

Different cosmologies are possible, mainly depending on whether
the  compositeness phase transition happened before, after or during  
cosmological inflation with Hubble constant $H_{\rm infl}$ 
driven by a vacuum energy $V_{\rm infl}$.
Assuming that inflation starts from a  cooling thermal bath with $g_*$ degrees of freedom, 
it begins at   temperature $T_{\rm infl}$ given by
\begin{equation}\label{eq:TinflH}
  \frac{g_* \pi^2 T^4_{\rm infl}}{30} =  V_{\rm infl} = \frac{3H_{\rm infl}^2M_{\rm Pl}^2}{8\pi}
\end{equation}
and ends giving a reheating temperature 
$T_{\rm RH}\approx T_{\rm infl} \min(1,\Gamma_{\rm infl}/H_{\rm infl})^{1/2}$
smaller than $T_{\rm infl} $ if 
the inflaton decay width $\Gamma_{\rm infl}$ is smaller than $H_{\rm infl}$.
The three main cases correspond to
\begin{enumerate}
\item  inflation after the compositeness phase transition if $T_{\rm RH} \le  T_{\rm infl} < f$;
\item inflation before  the compositeness phase transition if $f < T_{\rm RH} \le T_{\rm infl}$;
\item inflation during  the compositeness phase transition if $ T_{\rm RH}< f < T_{\rm infl}$.
\end{enumerate}
A case-by-case discussion would be lengthy.  We prefer to highlight
the new features that arise at the 
compositeness  phase transition, and that can be specialised to the various cases.

%\subsection{Inflation after the composite Higgs phase transition}
%The deepest minimum wins, and inflation does nothing.

\subsection{Compositness phase transition}
What happens at $T \lesssim f$ can be described by the effective field theory.
The Universe  randomly splits into domains of the various vacua
forming domains with typical size $R_0$ separated by domain walls.

The size $R_0$ is an important quantity which depends on details of the strong phase
transition that leads to the appearance of composite scalars.
It can be described by a QCD-like $\sigma$ field with mass $M_\sigma(T)$.
Regions have characteristic size $R_0 \sim 1/M_\sigma$.
This is not necessarily  microscopic:
$\sigma$ is massless at the critical temperature
(when global symmetry becomes broken) if the phase transition is of second order. 
The size of domains is then
limited by the time variation of the cosmological temperature.
Adapting the Kibble-Zurek computation~\cite{Kibble:1976sj,Zurek:1985qw} 
(see also~\cite{0905.1720}) to a  generic Hubble constant $H$ at the phase transition,
we find that the size of bubbles
\beq \label{eq:R0}
R_0 \sim \min \[ \frac{1}{f} \(\frac{f}{H}\)^{p},
\frac{1}{H}\],\qquad p= \frac{\nu}{1+\mu}
\eeq
depends on critical exponents $\nu$, $\mu$ 
that describe how the correlation length $\xi$
and
the relaxation time $\tau$  formally diverge close to the phase transition 
at temperature $T_{\rm cr}$ and time $t_{\rm cr}$:
\be 
\xi(t) \sim \frac{1}{f} \( \frac{T - T_{\rm cr}}{T_{\rm cr}}\)^{-\nu}\; \qquad \tau(t) \sim \frac{1}{f} \( \frac{T - T_{\rm cr}}{T_{\rm cr}}\)^{-\mu}.
\ee
Sufficiently far from the phase transition the relaxation time is microscopic, 
so that the system evolves by a sequence of equilibrium states. 
Because of the cooling due to the Hubble expansion  the phase transition is crossed at a finite rate. 
At a time $t_{\rm cr}-\tau$ close to the phase transition correlations freeze and determine the correlation length
\be \label{eq:xiKZ}
R_0 \approx \xi(t_{\rm cr}-\tau) \sim \frac{1}{f} \( - f \, T \, \frac{d t}{d T} \)^{p} = \frac{1}{f} \(\frac{f}{H} \)^p
\ee
having used $T \propto 1/a$ and thereby $dT/dt = -HT$.
%$T \propto e^{- H_I t}$ during inflation. 
This gives the first term in eq.\eq{R0}.
Due to lack of causal contact, $R_0$ must be below the Hubble scale $1/H$.
% A bubble with maximal $R_0 \circa{>} 1/H$ is a black hole.  True only at finite temperature

A first-order phase transition and a microscopic $R_0$ is obtained for $p \to 0$. 
%\(\frac{\bp}{T_{\rm cr}}\)^{\nu/(1+\mu)}\eeq
For a second-order transition $\nu = 1/(2- \gamma)$ where
$\gamma$ is the anomalous dimension of the squared mass of the order parameter. 
%$\mu$ is the critical exponent of the relaxation time, equal to $\mu \approx \nu$. 
In the `classical' Ginzburg-Landau limit $\nu = \mu =1/2$,
such that $p=1/3$.
In reasonable models $p<1$, such that  $R_0 \ll 1/H$ whenever $H \ll f$,
in particular during a thermal phase with $H \sim T^2/M_{\rm Pl}$.

\subsubsection*{Bubbles with $R_0 \ll 1/H$: microscopic black holes}
We next study how domains with typical size $R_0$ evolve.
We here consider the case of microscopic domains, $R_0 \ll 1/H$:
their evolution can be studied neglecting cosmology.
Assuming, for simplicity, a spherical bubble of false vacuum with 
thin-wall  with surface tension $\sigma \sim f^3$, the time evolution of
its radius $R$  is dictated by the conservation of
its mass/energy (see e.g.~\cite{Blau:1986cw,1512.01819})
\beq  M = 4\pi R^2 \sigma \sqrt{1 + \dot R^2% -H^2_{\rm in} R^2
} +\frac{4\pi R^3}{3} [\rho_{\rm in}-\rho_{\rm out} - 6\pi G{\sigma^2}].\eeq
The first term combines the surface and kinetic energy
(we neglect an extra term relevant on cosmological scales);
the second term is the mass excess, the
latter term is the gravitational energy of the wall, where $G=1/M_{\rm Pl}^2$ is the Newton constant.
Imposing $\dot M=0$ gives the classical equation of motion:
the deeper vacuum  expands into the false vacua
because vacuum energies have negative pressure.
Unless the energy difference is very small
(the special case of quasi-degenerate vacua will be considered in section~\ref{quasideg})
a bubble of false vacuum shrinks with velocity $\dot R $ which soon
becomes relativistic, and thereby on a time-scale much smaller than $H$.

The bubble can either disappear or form a black hole,
if its Schwarzschild radius $R_S\equiv 2 G M$ is larger than
the fundamental scale $1/f$, assuming that the
vacuum decay rate is negligible, and that
walls shrink loosing negligible energy to matter in the plasma, 
such that all the initial energy $M$ remains constant, becoming kinetic energy of walls.
In such a case a black hole forms when 
its radius $R$ becomes smaller than $R_S$.
In conclusion black holes are formed for 
%\LDL{[come ottenete questa eq.? assumendo $M \sim R_0^3 f^4$? Perche' non 
%$M \sim R_0^2 f^3$ da eq.~(20)?]  AS: $R_0 f\gg 1$}
\beq  R_0 \circa{>}(M_{\rm Pl}/f)^{2/3}/f \gg 1/f\eeq
with mass $M \circa{>}M_{\rm Pl}^2/f$.
Ignoring accretion, such black hole quickly evaporate  
in a time $t_{\rm ev} \sim G^2 M^3$ emitting Hawking radiation with temperature
 $T \sim 1/R_S$.

In conclusion, small (sub-horizon) bubbles of false vacuum just disappear, and the Universe gets
filled by the true vacuum.
An acceptable cosmology is obtained when the potential parameter 
$X = - V_1 / 8 V_4$ is positive {(cf.~fig.~\ref{fig:V0})},
as it means that the SM vacuum is the deepest vacuum.

\subsubsection*{Bubbles with $R_0 \circa{<} 1/H$: macroscopic black holes}
A more interesting situation happens if domains have horizon size $R_0 \sim 1/H$:
according to eq.\eq{R0} this only happens  for $H \sim f$.
Such a possibility is realised if the compositeness
phase transition occurs during inflation\footnote{See also~\cite{hep-ph/0106187} for a similar mechanism in the case of domain walls.}, 
assuming that it starts from a thermal phase with temperature
$T_{\rm infl}\sim (M_{\rm Pl} H)^{1/2}$ much larger than $f$
and proceeds with an exponential cooling, $T \approx T_{\rm infl} e^{-N}$ after $N$ $e$-folds of inflation.
Notice that $H \sim f$ needs either a low-scale inflation model
(e.g.\ $H \sim f\sim {\rm few}\times \TeV$) or a compositeness scale $f$ much above the weak scale, if $H \sim 10^{13}\GeV$ as in simplest inflationary models.
We actually assume that the inflationary Hubble rate 
is mildly smaller than $ f$, such that the dynamics of composite scalars is dominated by classical motion, 
rather than by inflationary fluctuations $\delta h\sim H/2\pi$.
This is also the situation that leads to black-hole
signals compatible with existing data, as we now discuss.

In such a case most bubbles have sub-horizon size $R_0\circa{<} 1/H$:
as discussed in the previous section they shrink forming
 small black holes that evaporate.\footnote{Black holes formed during inflation expands only mildly~\cite{gr-qc/9606052,1804.03462}, due to the change in $H$ as the inflaton rolls down its potential. We estimate their mass increase to be $\Delta M \sim H M^2/M_{\rm Pl}^2$, too small to make them macroscopic.}
 However, rare bubbles  have size $R_0$ larger than the correlation length $\xi$
 just because nearby regions can accidentally fluctuate in the same way
 with exponentially suppressed probability~\cite{percolation}
\be\label{eq:pR0}
\wp(R_0) \sim e^{- \alpha (R_0/\xi)^2}
\ee
for $\alpha\sim 1$.

Another effect helps some bubbles to reach Hubble size:
a bubble formed during inflation at time $t_{\rm cr}$ with radius $R_0$  initially does
not shrink because of thermal friction. 
The friction pressure is $\sim \dot R T^4$, 
whereas the pressure to the energy difference between the minima is $\sim V_1$. 
Therefore, a bubble with initial size $R_0$ keeps inflating down to
$T_s \sim V_1^{1/4}$, growing to size 
$
R_s \sim R_0 \sfrac{f}{V_1^{1/4}} $.
If $R_s \circa{>} 1/H$, the bubble inflates following the de Sitter geometry.
At the end of inflation, it can reach a large cosmological size.
After inflation the true vacuum expands, and the bubble
shrinks to a macroscopic black hole.
%
%When the thermal friction becomes subdominant, the bubble evolves with physical size $R(t) = e^{H_I t} r(t)$:
%\be 
%R(t) = \frac{1}{H_I} + e^{H_I(t-t_c)} \( R_s - \frac{1}{H_I}\)
%\ee
%approximating the bubble wall as shrinking at the speed of light, i.e. with $ds = 0$, $d r <0$. Therefore, if $R_s<1/H_I$ the bubble shrinks completely in a microscopic time. Instead, if the bubble has exited the event horizone, i.e.~$R_s>1/H_I$, 
This happens with  non-negligible (but suppressed) probability if
\be \label{eq:Hsuf}
\frac{H}{f} \lesssim \(\frac{V_1^{1/4}}{f}\)^{\frac{1}{1-p}}
\ee
but not much smaller than this, with again the first-order phase transition case recovered for $p \to 0$. 
%For instance, if $\nu = \mu \simeq 2/3$ and $\sqrt[4]{V_1/f^4} \sim 10^{-2}$ one gets $H_I/f \lesssim 5 \times 10^{-4}$. 
%If this happens, there is a sizeable probability $\mathscr{P}$ to have a bubble exiting the de Sitter event horizon, thus expanding exponentially and becoming of cosmological size. At the end of inflation they start shrinking and form a black hole if their size has not become too large, a cosmological domain filling the observable Universe otherwise.
%\subsubsection{Bubbles forming primordial giant black holes}
Let us estimate the mass and density of the population of such primordial black holes.
% formed by the collapse of the bubbles, by neglecting the energy loss due to thermal friction and accretion.
Denoting as $N_{\rm before} \sim \ln T_{\rm infl}/f$   the number of $e$-folds before 
the compositeness
phase transition, inflation lasts $N = N_{\rm before}+N_{\rm after}$ $e$-folds.
At the end of inflation, bubbles inflated to radius 
$R\sim e^{N_{\rm after}} /H$
%\sim  e^N f/(H_I T_{\rm infl})$
and have mass:
\beq
M \sim f^4 R^3 \sim e^{3N} \( \frac{f}{H}\)^3 \frac{f^4}{T_{\rm infl}^3} 
\sim e^{3(N-50)} M_\odot \(\frac{f}{10\TeV}\)^{5/2} \(\frac{0.01}{H/f}\)^{9/2}
%\gtrapprox e^{3N} \( \frac{f}{H_I}\)^3 \( \frac{T_{\rm RH}}{T_{\rm infl} }\)^3 f \notag\\
%&\simeq  e^{3(N-50)} \frac{f}{\unit{TeV}} \( \frac{f}{H_I}\)^3 \(\frac{T_{\rm RH}}{T_{\rm infl} }\)^3 \times 2.5 \times 10^{44} \, \unit{g}.s
\eeq
having assumed eq.\eq{TinflH}.
%\xxx{AS: che fine ha fatto il $3$ in $3N$?  PerchÂ non usi il legame fra $H$ e $T_{\rm infl}$?
%Sotto non ho ancora guardato} \DT{Typo, aggiustato. Nel conto non c'era typo. Non uso il legame per la mia domanda subito prima della (19). Parliamone?}
%where in the inequality we have used $f > T_{\rm RH}$. We stress again that this should only be considered as a rough estimate, since we have neglected the energy losses during the shrinking of the bubble. As a reference, the largest black holes observed so far have masses around $4 \times 10^{43} \, \unit{g}$. 
Ignoring accretion, black holes make
a fraction $ \sim \wp$ of the inflationary energy density,
such that their density is below the present dark matter density for
$\wp  \circa{<} T_{\rm eq}/T_{\rm RH}  \sim 10^{-13}  (10\TeV/T_{\rm RH})$ where
$T_{\rm eq} \sim0.7\eV$ is the temperature at matter/radiation equality,
and $\wp$ is given in eq.\eq{pR0}
and depends exponentially on model parameters.

%
%The average distance between the bubbles/black holes at the end of inflation is 
%\be 
%d \sim \frac{1}{H_I \mathscr{P}^{1/3}} \, e^N \, \frac{f}{T_{\rm infl}}
%\ee 
%which gives a density
%\be  
%\rho(T_{\rm RH}) \sim \mathscr{P} f^4 \;.
%\ee
%Since the radiation density is $\rho_{\rm rad} \sim T_{\rm RH}^4 < f^4$, and the black holes red-shift as matter, the probability $\mathscr{P}$ has to be rather small in order to not overclose the Universe, for instance $\mathscr{P} \lessapprox 10^{-12}$ for $f \sim \unit{TeV}$. If the Hubble rate during inflation is larger than as given by~\eqref{eq:Hsuf} (or even very close to the upper bound), the probability $\mathscr{P}$ is not suppressed and this scenario is excluded. 
%
%Finally, the average distance today between the black holes is
%\be 
%d_0 \simeq d \, \frac{T_{\rm RH}}{T_0} \sim e^{N-50} \, \frac{f}{H_I} \, \frac{T_{\rm RH}}{T_{\rm infl}} \,  \( \frac{10^{-12}}{\mathscr{P}}\)^{1/3} \times \unit[1.4]{Mpc}
%\ee
%having assumed no significant entropy production after reheating.

\subsection{(Quasi)degenerate minima and domain walls}\label{quasideg}
In the previous discussion we assumed that composite scalars have a potential with non-degenerate minima, 
finding that the deeper minimum expands into the false vacua.
As discussed in {sections~\ref{models} and \ref{PG}} some composite Higgs models
(such as the minimal {$\SO(5)/\SO(4)$} model of~\cite{hep-ph/0612048} {and the next-to-minimal 
$\SO(6)/\SO(5)$ model of~\cite{Gripaios:2009pe}} with fermions in the fundamental representation) predict degenerate vacua, 
which corresponds to the parameter $X=0$ in the notation of fig.~\ref{fig:V0}.
We here explore what happens in this situation.

The compositeness phase transition leads to
domain walls  similar to the $\mathbb{Z}_2$-walls discussed in \cite{Vilenkin:1984ib,Vilenkin:2000jqa}. 
%(see also \cite{Rai:1992xw}). 
Since the system is above the percolation threshold there is a single domain wall per horizon
much larger than the correlation length of the phase transition~\cite{Vilenkin:1981zs}
while the number density of finite closed walls is exponentially suppressed.  
The evolution of the wall is governed by the pressure $p_T$ due to 
wall tension (which tends to minimise the wall area)
and by the frictional pressure $p_F$ with the surrounding medium. 
%\xxx{AS: e l'energia gravitazionale $G\sigma^2$?Il  dominio  grosso, ma il raggio di curvatura quando si forma  ~ xi. Quindi ai fini dei buchi neri conta come bolle piccole (in altre parole,  una bolla grossa che ha tantissime bolle piccole dentro). Per la cosa dei domain wall invece conta come uno (Perch in assenza di splitting fra i minimi l'energia di superficie lo fa stirare).}
The former is $p_T \sim \sigma / R$, 
where $\sigma \sim f^3$ is the wall tension 
%(energy per unit area) 
and $R$ is the radius of curvature  
of the wall structure, and the latter is $p_F \sim v T^4$, where $v$ is the wall velocity and $T$ the temperature of the 
surrounding medium. At very early times the walls are over-damped and their velocity is determined by the balance of the two forces $p_T \sim p_F$. 
Assuming radiation domination ($T^4 \sim \rho \sim M_{\rm Pl}^2 / t^2$) 
this yields $v \sim \sigma t^2 / (R M_{\rm Pl}^2)$. The region over which the wall is smoothed out at time 
$t$ is $v t$ and hence it grows with time as $R(t)^2 \sim \sigma t^3 / M_{\rm Pl}^2$. 
The contribution of the walls to the energy density is then 
$\rho_{\rm wall} \sim \sigma / R(t) \sim \sigma^{1/2} M_{\rm Pl} / t^{3/2}$, 
which normalized to the critical energy density $\rho_{\rm cr} \sim M_{\rm Pl}^2 / t^2$ 
yields $\Omega_{\rm wall} = \rho_{\rm wall} / \rho_{\rm cr} \sim (t \sigma)^{1/2} / M_{\rm Pl}$. 
The domain wall start  dominating the energy density at 
$t_{\rm wall} \sim M_{\rm Pl}^2 / \sigma \sim 100 \ \text{s}\times  \( 10 \ \text{TeV} / f \)^3 $,
in contradiction with observations.

These considerations imply that the minimal model of~\cite{hep-ph/0612048} based on the coset ${\rm SO}(5)/{\rm SO}(4)$ with fermions in the $\rm 5$  of {\rm SO}(5), {as well as 
the next-to-minimal ${\rm SO}(6)/{\rm SO}(5)$ model of \cite{Gripaios:2009pe} with fermions in the $\rm 6$, are  
excluded by their} unacceptable cosmology, unless inflation occurs at very low scale, after the composite phase transition.

\medskip

A possible way out is a small difference $V(h=0) - V(h=\pi f) = 2 V_1$ between the energy 
densities of the two vacua, 
so that the deepest vacuum dominates and the wall disappears. 
The corresponding pressure on the wall is $2V_1$, which must
overcome the wall tension before the domain wall  
starts to dominate the energy density. 
Hence, one obtains the condition \cite{Vilenkin:1981zs,Gelmini:1988sf,Rai:1992xw}\footnote{This 
simple estimate is supported by numerical simulations of 
the domain wall network evolution \cite{Larsson:1996sp}.}
\beq 
\label{eq:Cbound1}
V_1 > p_T \sim \frac{\sigma}{R} > \frac{\sigma^2}{M_{\rm Pl}^2} \sim f^4 \( \frac{f}{M_{\rm Pl}} \)^2
\qquad \Rightarrow \qquad
\left| \frac{V_1}{f^4} \right| > \( \frac{f}{M_{\rm Pl}} \)^2 > 10^{-30} \( \frac{f}{10 \ \text{TeV}} \)^2  .
\eeq
A parametrically different bound is obtained  
requiring that the domain wall disappears before nucleosynthesis 
($R_H \sim M_{\rm Pl} / T^2_{\rm BBN}$, with $T_{\rm BBN} \sim 1$ MeV):
\beq
\label{eq:Cbound2}
V_1 > \frac{\sigma}{R_H} \sim \frac{f^3 T^2_{\rm BBN}}{M_{\rm Pl}}
\qquad \Rightarrow \qquad
\left| \frac{V_1}{f^4} \right| >  \frac{T^2_{\rm BBN}}{f M_{\rm Pl}} > 10^{-29}  \frac{10 \ \text{TeV}}{f}.
\eeq
In conclusion,  a small 
$| V_1 / f^4 |$ 
%of the order of $10^{-20}$ 
is enough to remove the domain wall issue. 
In both the minimal \cite{hep-ph/0612048} and next-to-minimal \cite{Gripaios:2009pe} 
composite Higgs models non-degeneracy can be achieved by introducing almost-decoupled fermions in the spinorial representation of SO(5) and SO(6), respectively.

%at the expense of the original minimality of the model. 
%\LDL{[Riusciamo a dirlo meglio? Per minimal model comunemente 
%si intende Contino2004 che non ha domain wall issue]}
%If $f\circa{>}10^{15}\GeV$, a large enough $V_1$ can be provided by the gravitational instantons of section~\ref{Gravity}. 

%\begin{figure}[ht]
%$$\includegraphics[width=0.6 \textwidth]{figs/C_boundDW}$$
%\caption{\em\label{fig:C_boundDW} 
%Bound on $\left| V_1 / f^4 \right|$ from condition eq.\eq{Cbound1} (blue) and eq.\eq{Cbound2} (yellow).}
%\end{figure}

%\LDL{If you think the plot is too trivial we can write (using $M_{\rm Pl} = 10^{19}$ GeV and 
%$T_{\rm BBN} = 1$ MeV):
%\be 
%\left| \frac{V_1}{f^4} \right| > 10^{-30} \( \frac{f}{10 \ \text{TeV}} \)^2 
%\ee
%\be 
%\left| \frac{V_1}{f^4} \right| > 10^{-29} \( \frac{10 \ \text{TeV}}{f} \) 
%\ee
%}

\subsection{Bubbles filling the observable Universe}
As discussed previously, false vacuum bubbles can become exponentially large
when (after inflation) true vacuum expands in them:
an observer outside sees a black hole remnant.
According to general relativity the interior is not affected by 
classical expansion of the true vacuum, and forms a baby Universe~\cite{Blau:1986cw}.
Another possibility is then that our observable Universe was inside a false vacuum bubble.
This is possible provided that quantum or thermal tunnelling (ignored so far) towards the true vacuum
is fast enough.
The answer is model-dependent.

%Any one of the local minima could have filled our horizon through inflation.
%If the pre-inflation history is thermal, this needs that inflation happens just after the composite Higgs phase transition.
%
%We could live in the unstable minimum if it is enough long-lived
%(or if it is dS and enough short-lived that excessive super-cooling is avoided).

\begin{figure}[t]
$$\includegraphics[width=0.5 \textwidth]{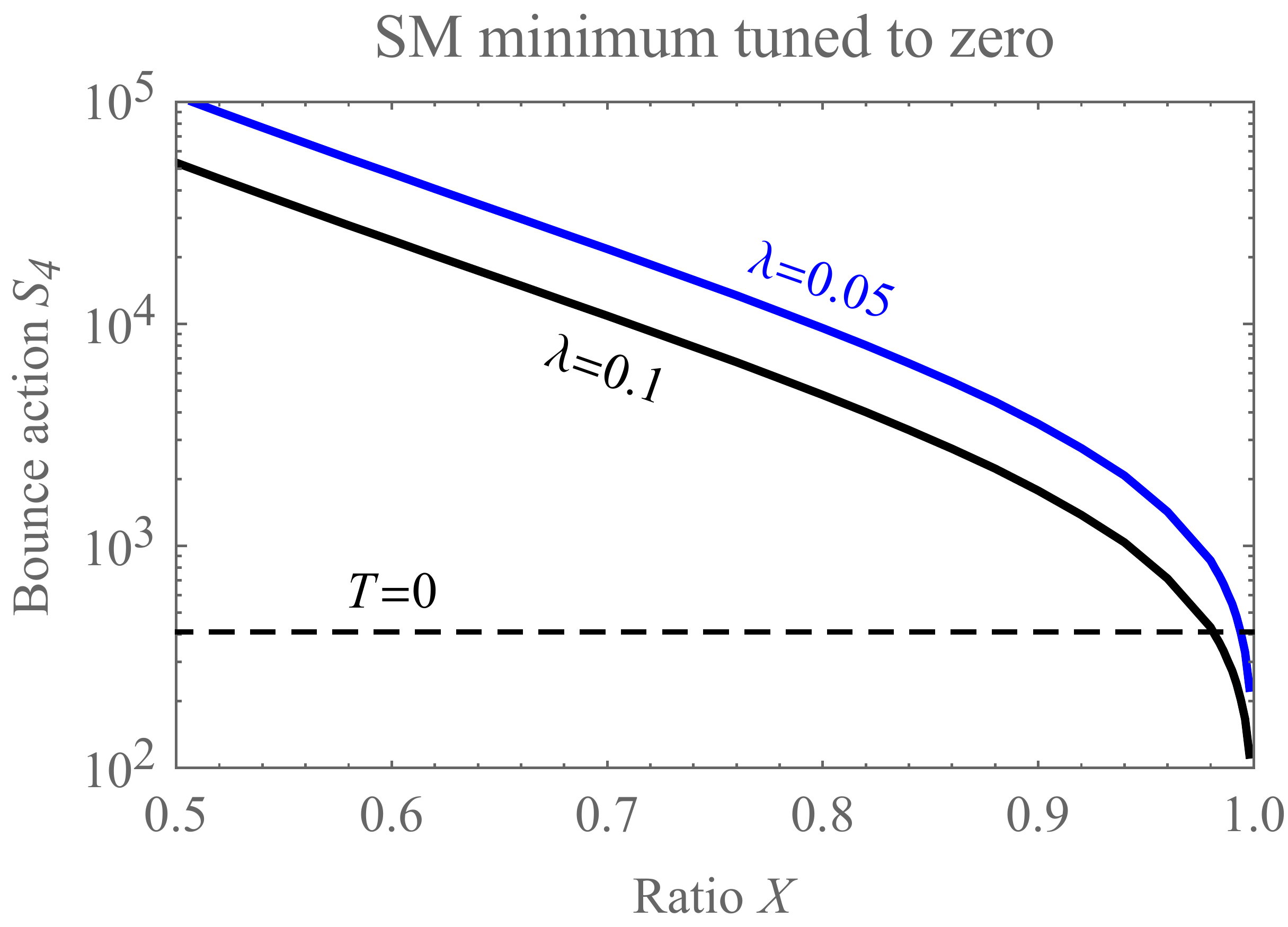}$$
\caption{\em\label{fig:bounce0T} Action of the bounce for quantum vacuum decay at zero temperature
as function of the free parameter $X$ defined in eq.\eq{R}. 
Cosmologically fast rates are obtained for $S_4\circa{<}400$.}
\end{figure}

\subsubsection*{Quantum vacuum decay}
The space-time density of vacuum decay probability is approximated by $dp/d^4x\sim e^{-S_4}/f^4$,
where $S_4$ is the Euclidean action of the bounce, a classical solution 
that interpolates between the minima~\cite{Coleman:1977py}.
This extends straightforwardly to a Goldstone boson as long as we are careful to canonically normalize the kinetic term.

Let us consider the composite Higgs.
We can conveniently work in the basis where $h$ has a canonical kinetic term: the Higgs
potential of eq.\eq{Vcos} depends on $\cos nh/f$, so that it is  convenient  to rescale the space-time coordinates $x_\mu$ and
$h(x_\mu)= f\, \tilde{h}(\tilde x_\mu/\sqrt{\lambda} f)$ 
to dimension-less variables $\tilde h$ and $\tilde x_\mu$.
The Euclidean action becomes 
\beq\label{eq:VABCtilde}
S_4 =\frac{1}{\lambda}\int d^4\tilde x \bigg[\frac{(\tilde\partial_\mu \tilde h)^2}{2} +\sum_{n=0}^\infty  \frac{V_n}{\lambda f^4} \cos {n \tilde{h}}\bigg] \equiv \frac{F(X)}{\lambda}\eeq
where 
$\lambda$ is a free parameter that can be used to get rid of the overall scale in $V$, such that the Higgs
potential with tuned $v \ll f$ only depends on $X=-V_1/8V_4$, as defined in eq.\eq{R}.
We choose $\lambda$ equal to the Higgs quartic coupling of eq.\eq{VSM}.
%\beq
%S =\frac{1}{B}\int d^4\tilde x \bigg[\frac{(\tilde\partial_\mu \tilde h)^2}{2} + \bigg( \frac{A}{B} \sin^2 \tilde{h} +  \sin^4\tilde{h} - \frac{C}{B} \cos\tilde{h} \bigg)\bigg] \equiv \frac{F(C/B)}{B}.\eeq
The ${\rm O}(4)$-invariant bounce solution is computed numerically
and plugged into eq.~\eqref{eq:VABCtilde} obtaining the bounce action $S_4$ plotted in fig.~\ref{fig:bounce0T}.
Given that $ \lambda \sim 0.1$,  vacuum decay is cosmologically
fast for $0.98\circa{<}X \le 1$, which corresponds to a small enough potential barrier. 
In this restricted range of $X$ we can live in a large bubble of
false vacuum that decayed fast enough to the SM vacuum.

A metastable SM vacuum ($X<0$) 
can also be long-lived enough to be cosmologically acceptable.
%\LDL{[Su diciamo: ``An acceptable cosmology is obtained when the potential parameter 
%$X = - V_1 / 8 V_4$ is positive {(cf.~fig.~\ref{fig:V0})},
%as it means that the SM vacuum is the deepest vacuum.'']} \DT{Qua ci riferiamo alla transizione di fase avvenuta durante l'inflazione. Sopra transizione prima o dopo. Se si vuole si può specificare esplicitamente in entrambi i casi.}

\begin{figure}[t]
$$\includegraphics[width=0.5 \textwidth]{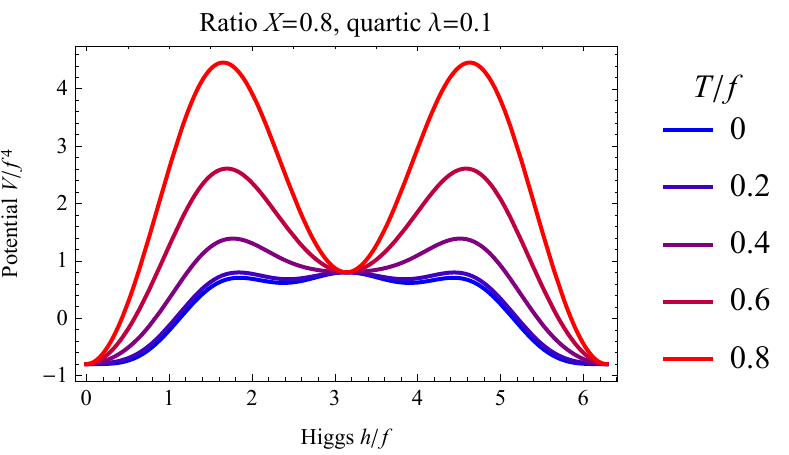} \quad \includegraphics[width=0.5 \textwidth]{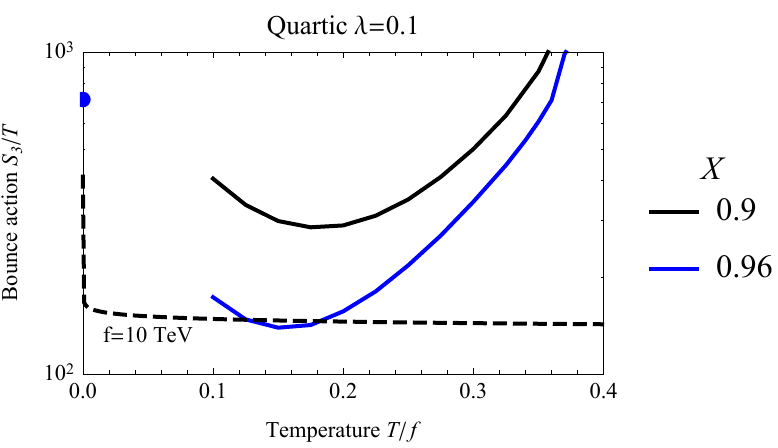}$$
\caption{\em\label{fig:case1} We consider the model of eq.\eq{case1}.
{\bf Left}: thermal potential. 
{\bf Right}: $S_3/T$ action of the ${\rm O}(3)$ bounce (full curves).
The dot shows the
 action of the ${\rm O}(4)$ bounce relevant at $T \ll f$.
Below the dashed line tunnelling is cosmologically fast during radiation domination.}
\end{figure}

\begin{figure}[t]
$$\includegraphics[width=0.5 \textwidth]{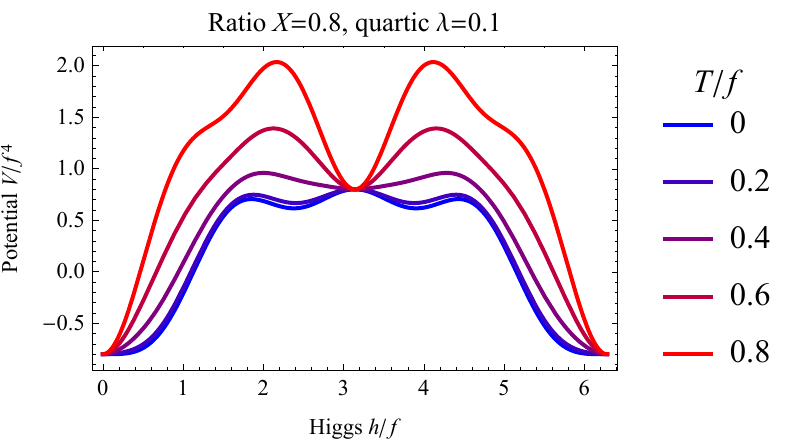} \quad \includegraphics[width=0.5 \textwidth]{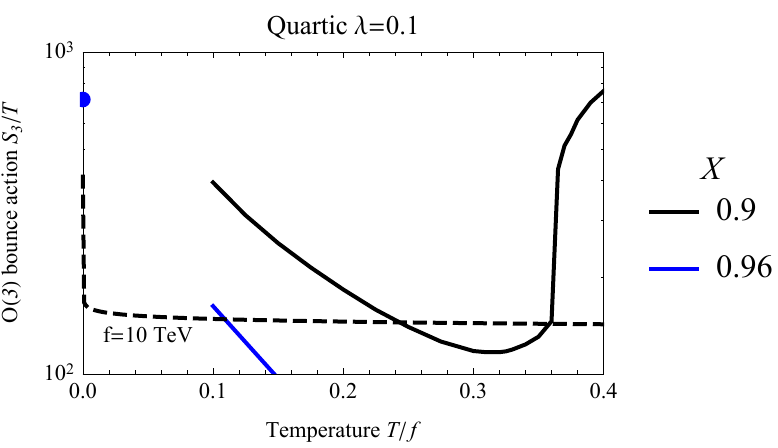}$$
\caption{\em\label{fig:case2} Same as in fig.~\ref{fig:case1}, for the model of eq.\eq{case2}.}
\end{figure}

%\begin{figure}[t]
%$$\includegraphics[width=0.5 \textwidth]{figs/caso3_CsuB08} \quad \includegraphics[width=0.5 \textwidth]{figs/bounce_caso3}$$
%\caption{\em\label{fig:case3} Thermal potential, case 3. Action of the $O(3)$ bounce $S_3/T$ in Case 1. We also show the action $S_4$ of the ${\rm O}(4)$-invariant bounce at $T=0$ and the dashed line denotes the critical action to have tunnelling assuming radiation domination.}
%\end{figure}
%
%\begin{figure}[t]
%$$\includegraphics[width=0.5 \textwidth]{figs/caso4_CsuB08}$$
%\caption{\em\label{fig:case4} Thermal potential, case 4.}
%\end{figure}

\subsubsection*{Thermal tunneling}
The rate for vacuum decay at finite temperature is computed  from the thermal potential,
finding a bounce solution with periodicity $1/ T$ in the Euclidean time direction.
Thereby the ${\rm O}(4)$-invariant bounce remains the dominant configuration at $T \ll f$,
while at $T \gg f$ ($T \gtrsim 0.1 f$ in composite Higgs) the dominant configuration is ${\rm O}(3)$-invariant and constant in Euclidean time,
with action $S_3/T$ (see e.g.~\cite{hep-ph/9901312}):
\beq \frac{S_3}{T}  =\frac{f}{T \lambda^{1/2}} \int d^3\tilde x \bigg[\frac{(\tilde\partial_\mu \tilde h)^2}{2} +\sum_{n=0}^\infty  \frac{V_n}{\lambda f^4} \cos {n \tilde{h}} + \frac{V_T}{\lambda f^4}\bigg] \equiv \frac{f\, F_T(X;T)}{T \lambda^{1/2}}
\eeq
where $V_T$ is the thermal contribution to the potential. Depending on the shape of the potential, more complicated configurations can be relevant at intermediate temperatures~\cite{oldTM}.

\medskip

Usually a faster thermal decay arises at temperatures above the mass scale
in the potential.  Furthermore, thermal corrections to the potential usually 
tend to remove the false vacuum.
As discussed at the beginning of section~\ref{cosmo},
the case of a pseudo-Goldstone boson is special:
the thermal potential given in eq.\eq{VT} 
can have multiple minima, separated by increasing barriers at large $T$.
Details are model-dependent and fig.\fig{case1} shows an example in this direction.
Considering models~\cite{hep-ph/0412089,1607.01659}  where
\be \label{eq:case1}
M_W^2 = \frac{g_2^2 f^2}{4} \sin^2\frac{h}{f} ,\qquad M^2_t = \frac{y_t^2 f^2}{2}  \sin^2\frac{h}{f}  
\ee
we see that thermal tunneling can become cosmologically fast in a
narrow range of temperatures $T \sim 0.15 f$ where our computation is trustable. This results from the competition of two factors: as usual, thermal tunnelling tends to become faster as the temperature increases, 
however in our case the thermal barriers slow the tunnelling growth too. A faster tunnelling rate is obtained, for example, in a model with~\cite{hep-ph/0612048}
\be \label{eq:case2}
M_W^2 = \frac{g_2^2 f^2}{4} \sin^2\frac{h}{f} ,\qquad M^2_t = \frac{y_t^2 f^2}{8}  \sin^2\frac{2 h}{f}  
\ee
as shown in fig.~\ref{fig:case2}. Moreover, in this case for $X \gtrsim 0.95$ the second minimum disappears in a small temperature interval, and the field reaches the true vacuum by rolling, rather than tunnelling.

%This approximation is valid as long as $T/f \gg \mathcal{O}(1) \sqrt{B} \simeq 0.08$.
%For the thermal potential, we distinguish two cases, as described in Sec.~\ref{sec:VYukawa}

%\paragraph{Case 1: $\sin^2(h/f)$}
%In the first case, the function $\phi$ in \eqref{eq:VYukawaT} is $\phi^2(h/f) = \sin^2(h/f)$. Taking also into account the gauge contribution  only $A$ receives a thermal correction $A_T$:
%\begin{equation}
%A_T = \frac{2 m_W^2 + m_Z^2 + 2 m_t^2}{8 v^2} \, \frac{T^2}{f^2}
%\end{equation}
%For fixed values of $C/B$ and the relative temperature $\tilde{T} \equiv T/f$, the ratio $A_T/B$ is determined by fitting the quartic $\lambda$. The results for $C/B = 0.98$ are shown in Fig.~\ref{fig:case1}.
%
%*******
%
%add comments, details; unique features of the thermal tunnelling in this case, slower than $T=0$.
%
%*******

%*******
%
%add comments, details; unique features of the thermal tunnelling in this case, competition between $O(3)$ faster than $O(4)$ and thermal barriers growing with $T$. In the example plot  tunnelling is active for finite time: cosmological consequences?

The false vacuum can even become the true vacuum at finite temperature. 
This happens, for instance, in the presence of new particles that become massless
e.g.\ at $h=f\pi$ but not at $h=0$. 
Another possibility is an 
unsuppressed wave-function contribution $Z_U \sim \cos (h/f)$ 
in the first term of eq.~\eqref{eq:VYukawaT02}, 
%as e.g. in the model of [8], 
such that at finite temperature the $h = f\pi$ minimum is favoured.
This can in principle realise the idea of electro-weak baryogenesis above the weak 
scale~\cite{1807.07578,1807.08770,1811.11740} (see also~\cite{hep-ph/0409070})
without requiring a large number of fields. 

Consider, for example, a model where $M_t=0$ but $M_W\neq 0$ at the $T=0$ false vacuum at  $h \sim \pi f$. 
If at high temperature $T\sim f$ this becomes the deepest minimum, 
the compositeness phase transition will populate it (with the microscopic bubbles of $h \simeq 0$ shrinking fast). 
At $h \simeq \pi f$ the electro-weak gauge bosons have large masses $\sim f$, so that the sphalerons are suppressed and electro-weak baryogenesis can take place, if the composite phase transition is first-order and sufficient CP violation is present. At low temperatures the SM minimum $h = v \ll f$ becomes the global minimum and populates the Universe, either by rolling or tunnelling. 
If this happens at a temperature $T \lesssim \unit[130]{GeV}$ electro-weak sphalerons remain frozen and the baryon asymmetry is not washed out.
As argued in~\cite{1807.07578,1807.08770,1811.11740}, the advantage of having electro-weak baryogenesis well above the weak scale is that the required CP violation is much less constrained than in the increasingly challenged models at the weak scale. 

However, we find it difficult to realize this scenario quantitatively in the models considered in this work: the Higgs field must be at the $h = \pi f$ minimum until temperatures $T \lesssim \unit[130]{GeV}$, i.e. $T/f \lesssim 0.04$ for the phenomenologically acceptable values $f \gtrsim \unit[3]{TeV}$. Instead, for the models considered here 
we find that, according to parameters, tunnelling either occurs for $T/f \gtrsim 0.1$ or does not occur altogether (see fig.s~\ref{fig:bounce0T}--\ref{fig:case2}). Exploring this possibility in more complicated models, possibly involving more scalar fields, goes beyond the scope of this work and might be done elsewhere.

\section{Conclusions}\label{concl}
Pseudo-Goldstone bosons, such as the composite scalars arising from new strong dynamics,
can be described by low-energy effective theories where the scalar fields form a coset with non-trivial topology.
We found that their scalar potential often admits multiple minima.
Selecting one scalar, its field space is a circle along which the kinetic term can be made canonical:
in this basis its interactions with SM particles and its potential  contain `trigonometric'
terms that go beyond the polynomial terms of low-energy renormalizable theories.

In section~\ref{models} we considered the potential of a composite Higgs boson:
for phenomenological reasons the trigonometric potential of eq.\eq{Vcos} 
cannot be dominated by the term with largest period,
such that terms with smaller periodicities can give rise to multiple minima.
We provided simple expressions for the quantum and thermal corrections to the potential
generated by low-energy Higgs interactions with the $W,Z$ bosons and with the top quark,
parameterized in eq.~(\ref{sys:MWh}) and eq.\eq{Mtper} by their model-dependent periodicities.
The presence of multiple points in field space where
$M_{W,Z}$ and/or $M_t$ vanish is the reason for the presence of multiple minima, even at finite temperature.
No extra scalars are present in the minimal composite Higgs model based on the $\rm SO(5)/SO(4)$:
the potential has one SM minimum degenerate with an anti-SM minimum,
unless spinorial representations are introduced.

In section~\ref{PG} we considered the potential along the full coset in different models,
relevant for composite Higgs and/or composite Dark Matter.
We found a variety of behaviours.
In the composite Higgs model based on SO(6)/SO(5) the two minima of the Higgs potential
get connected by another scalar: its potential can have or not have a barrier, as exemplified in fig.\fig{Palla}.
We also considered models where an $\SU(N_c)$ new strong gauge interaction and $N_F$ flavours of `dark quarks'
leads to a coset with $\SU(N_F)$ topology, equal to $S^3$ for $N_F=2$ and to $S^3\times S^5$ for $N_F=3$.
When dark quarks are charged under the SM gauge group, the gauge contribution to the pseudo-Goldstone bosons potential
has minima at the $N_F$ centers of $\SU(N_F)$ and along singlet directions.
Dark quark masses and/or the $\theta$ angle of the strong gauge group and/or Yukawa interactions
can break their degeneracy. Various cases have been explicitly computed in appendix~\ref{QCDapp}, finding a variety of behaviours which
includes multiple local minima.

In section~\ref{cosmo} we explored the cosmological consequences.
An interesting feature of pseudo-Goldstone bosons potentials is that
multiple minima tend to remain present at finite temperature and with higher barrier,
because thermal corrections to the potential are generated by interactions with particles 
(the SM vector bosons, the top quark, etc.) that are light at multiple point in the coset field space.
Degenerate minima lead to problematic domain wall issues, unless the degeneracy can be lifted. In general the minima are
not degenerate: the deeper minimum expands into the false vacua.  Therefore, if the phase transition occurs strictly before or after inflation, an acceptable cosmology is obtained only if the minimum at $h \simeq 0$ is the global minimum of the potential, thus restricting  parameters of the model  potentially coming from UV physics and inaccessible otherwise.
The shrinking of false-vacuum bubbles can leave black hole remnants, which are microscopic and evaporate quickly unless
the compositeness phase transition happens during inflation, leading to Hubble-sized domains
that inflate.  In such a case the true vacuum expands into the false vacuum only after inflation,
leaving supermassive macroscopic black holes.
A related different possibility is that we live inside a false vacuum bubble, which decays fast enough through thermal or quantum tunnelling to the true vacuum.  
In general, the $W,Z$ bosons can be massive inside the false vacuum:
this new kind of minima could have implications for electro-weak baryogenesis.

\subsubsection*{Acknowledgements}
This work was supported by the ERC grant NEO-NAT.
We thank Roberto Contino, Luigi Delle Rose, Ramona Gr\"{o}ber, Maxim Khlopov, Geraldine Servant,  Andrea Tesi and Nikolaos Tetradis for discussions.

\appendix

\small

\section{On the (in)equivalence of Higgs configurations}\label{ineq}
We here explicitly show that $h=f\pi$ is not equivalent to $h=0$ considering the composite Higgs model with the minimal coset, $\SO(5)/\SO(4)$,
which is a higher-dimensional sphere.
%The potential can have two minima at these two opposite poles.

%The two minima are related by a global transformation in the coset $SO(5)/SO(4)$, since these correspond to shifts of $h$. 
The two points would be the same point if a gauge transformation existed, that connected them. 
However, this is not the case. 
Gauge transformations are embedded in the unbroken group $\mathscr{H}$; therefore, their generators are not in the coset $\mathscr{G}/\mathscr{H}$ that connects the two minima. As a consequence, the two minima are connected by a global but not local transformation, so that they are two distinct points. 
To see this argument more explicitly, let us consider the broken generators in the coset $\SO(5)/\SO(4)$ 
\begin{equation}
(T_a)_{ij} = - \frac{i}{\sqrt{2}} \big[ \delta_i^a \delta_j^5 - \delta_j^a \delta_i^5 \big]
\end{equation}
with the alignment of the vev $\vec\Sigma_0 = \Sigma_0 (0 , 0 , 0 , 0 , 1 )$.
The Higgs boson is given by the excitations along the coset, i.e.~
\begin{equation}\label{eq:HinSO5}
\Sigma = \Sigma_0 \, e^{- i \sqrt{2} \, T_a h_a/f} =  (0, 0 , \sin h/f, 0 , \cos h/f ) 
\end{equation}
having exploited part of the gauge redundancy to align the doublet in the
3rd component of $\vec \Sigma$.
The $\textrm{SU(2)}_{L,R}$ generators, embedded in $\SO(4)$  are
\begin{align}
T^{L,R}_1 = -\frac{i}{2} \left( \begin{smallmatrix}
0 & 0 & 0 & \pm 1 & 0\\
0 & 0 & 1 & 0 & 0 \\
0 & -1 & 0 & 0 & 0 \\
\mp 1 & 0 & 0 & 0 & 0 \\
0 & 0 & 0 & 0 & 0 
\end{smallmatrix} \right), \quad 
T^{L,R}_2 = -\frac{i}{2} \left( \begin{smallmatrix}
0 & 0 & -1 & 0 & 0 \\
0 & 0 & 0 & \pm 1 & 0 \\
1 & 0 & 0 & 0 & 0 \\
0 & \mp 1 & 0 & 0 & 0 \\
0 & 0 & 0 & 0 & 0 
\end{smallmatrix} \right),
\quad 
T^{L,R}_3 = -\frac{i}{2} \left( \begin{smallmatrix}
0 & 1 & 0 & 0 & 0 \\
-1 & 0 & 0 & 0 & 0 \\
0 & 0 & 0 & \pm 1 & 0 \\
0 & 0 & \mp 1 & 0 & 0 \\
0 & 0 & 0 & 0 & 0 
\end{smallmatrix} \right).
\end{align}
Since they vanish on the 5th component, they cannot perform the shift $h \to h + \pi f$ in~\eqref{eq:HinSO5}. For instance, $\SU(2)_L$ gauge transformations of the Higgs field with gauge parameters $\vec{\alpha}$ give
\begin{align}
&\Sigma \, e^{- 2 i \pi \vec{\alpha} \cdot \vec{T}^L} = 
\begin{pmatrix} - \frac{\alpha_2 \sin(\pi \alpha)}{\alpha} \sin\big(\frac{h}{f}\big) &  \frac{\alpha_1 \sin(\pi \alpha)}{\alpha} \sin\big(\frac{h}{f}\big) & \cos{(\pi \alpha)} \sin\big(\frac{h}{f}\big) & - \frac{\alpha_3 \sin(\pi \alpha)}{\alpha} \sin\big(\frac{h}{f}\big) & \cos\big(\frac{h}{f}\big) \end{pmatrix}
\end{align}
with $\alpha \equiv |\vec{\alpha}|$, so that the 5th component is left unchanged. This also shows that, instead, the transformation $h \to - h$ is a gauge transformation with $\alpha = 1$. To summarize, $h/f$ has period $2 \pi$, with the gauge symmetry imposing that the potential is an even function of $h$.\footnote{This conclusion is also reached considering the simpler analogous case
$\mathscr{G}/\mathscr{H}=\SO(3)/\SO(2)$, closer to our geometrical intuition. 
One might worry that the $\mathbb{Z}_2$ appearing in the double-covering relation $\textrm{SO(3)} \sim \textrm{SU(2)}/\mathbb{Z}_2$ could identify  $h=0$ with $h=f\pi$. 
This is not the case.
The SO(3) manifold is the solid ball of radius $\pi$ in 3 dimensions,
with antipodal points identified. This is because any 3-dimensional rotation is uniquely determined by an axis and an angle $-\pi \leq \theta \leq \pi$, with the two rotations of $\pm \pi$ being the same:
this is the $\mathbb{Z}_2$ identification.
The two points $h=0$ and $h=f\pi$ are distinct: $h=0$ corresponds to the centre of the ball, whereas $h=\pi f$ corresponds to the two identified points on the boundary of the ball along a given direction. }
%Notice also that since $\mathrm{SO}(N)/\mathrm{SO}(N-1) \sim \mathcal{S}^{N-1}$, the two minima are the two distinct poles of the $N-1$-dimensional sphere that represents the coset manifold.

\section{Potential in QCD-like examples}\label{QCDapp}
In this appendix we discuss in detail potential of pseudo-Goldstone bosons in QCD-like theories.

\subsection{Multiple minima from the $\theta$ angle}\label{theta2}
In general to discuss the effect of the  $\theta$ angle it is convenient to include
the heavy $\eta'$ singlet in the effective low energy theory~\cite{Witten:1980sp,DiVecchia:2017xpu,Gaiotto:2017tne}.
The low energy Lagrangian is described by a unitary matrix ${\rm U}(N_F)$ matrix with the action  eq.~(\ref{lagrangianN}) supplemented by the anomaly term
\begin{equation}
V_{\rm anomaly}=-\frac{f^2}{16}\frac c N_c \bigg[\ln (\hbox{det}\ \mathscr{U})- \ln (\hbox{det}\ \mathscr{U}^\dagger)\bigg]^2 \;,
\end{equation}
such that $m_{\eta'}^2\approx 3 c/N_c$ becomes light at large $N_c$.
In view of the determinant, we can compute $V_{\rm anomaly}$ restricting to the diagonal ansatz 
\begin{equation}
\mathscr{U}=e^{-i \sfrac{\theta}{N_F}}\hbox{diag}\,(e^{i\phi_1},\dots,e^{i\phi_{N_F}})
\label{Uvev}
\end{equation} 
obtaining
\begin{equation}
V_{\rm mass}+V_{\rm anomaly}\approx \frac{f^2} 4 \left[ - 8 \sum  \mu_i^2 \cos \phi_i+ \frac c N (\sum_i \phi_i- \theta)^2\right]
\end{equation}
where $\mu_i^2= g_* f   M_{\Q_i}$. The extrema of the potential correspond to the solutions of Dashen equations
\begin{equation}
4  \mu_i^2  \sin\phi_i = \frac{c}{N_c} (\theta - \sum_i \phi_i) \,.
\end{equation}
In the limit of small masses these equations impose $ \sum_i \phi_i=\theta $, reducing to what written in
section~\ref{QCD}.
These equations admit multiple solutions, for certain range of masses.
%Consider first the case of 3 degenerate flavors, as in section~\ref{Q=3}.
%Assuming 3 equal phases one obtains the single equation
%\begin{equation}
%4 \mu^2 \sin\phi = \frac{a}{N} (\theta -3 \phi)
%\end{equation}
%Multiple solutions exist when $4 \mu^2 > m_\eta'^2/3$. Physically also $\mu^2<m_\eta'^2$ for
%the theory to make sense so that at most $O(1)$ extra solutions exist for moderate values of $N$. In the opposite regime $\mu^2 \ll m_\eta'^2$ the approximate
%solution is $\phi=\theta/3$. This solution however is not the lowest energy solution of the system for every $\theta$ because it is not $2\pi$ periodic. In order to find 
%the true vacuum we need to allow one of the three phases to be different. As we show next this is consistent with $SU(3)$ global symmetry due to the periodicities.
Assuming, for example, two degenerate flavors and a singlet ($\Q=\Q_L\oplus \Q_N$,  in section~\ref{partiallycomposite})
we look for a solution with two equal phases $\phi_2$ and a phase $\phi_1$. 
Dashen equations take the form
\begin{equation}
4 \mu_2^2 \sin\phi_2=4 \mu_1^2 \sin\phi_1  = \frac{c}{N} (\theta - 2 \phi_2-\phi_1)
\end{equation}
These equations can be solved numerically and lead to multiple vacua for $\mu_2^2 \le 2 \mu_1^2 \ll c/N$.
The solutions cross at $\theta=\pi$ where the energy is degenerate breaking  CP spontaneously as we now show.
An analytic approximation (equivalent to integrating out the $\eta'$) 
is obtained by noting that $\theta - 2 \phi_2-\phi_1 \approx 0$
implies
\begin{equation}
\mu_2^2 \sin \phi_2= \mu_1^2 \sin (\theta-2 \phi_2) \,,
\end{equation}
which leads to an algebraic equation for $\sin \phi_2$. For $\mu_1=\mu_2$ the solutions are
\begin{equation}
\left\{ 
\begin{split} 
&\phi_1= \frac {\theta}3- \frac{4\pi}3 n \,~~~~~~~~~~~~~{\rm and}~~~~~~~~~ \phi_2=\frac{\theta}3 +\frac {2\pi}3 n\\
&\phi_1= -\theta+ \pi (2n+1) \,~~~~~{\rm and}~~~~~~~~~ \phi_2=\theta - \pi (2n+1)
\end{split} \right.
\end{equation}
For $\mu_1\neq\mu_2$ the solution is simple for $\theta=0,\pi$:
\begin{equation}
\left\{ 
\begin{split} 
&\theta=0:~~~\sin \phi_2=0 \,~~~~~~~~~{\rm or}~~~~~~~~ \cos \phi_2=-\frac {\mu_2^2}{2\mu_1^2}\\
&\theta=\pi:~~~\sin \phi_2=0 \,~~~~~~~~~{\rm or}~~~~~~~~~ \cos \phi_2=\frac{\mu_2^2}{2\mu_1^2}
\end{split} \right. \,,
\end{equation}
where each solution corresponds to two physical points. Considering the mass matrix of the Goldstone bosons
\begin{eqnarray}
\label{pionmasses}
m^2_{\pi_3}&=&  4 M_{\Q_2} \cos \phi_2 g_* f \nonumber\\
m^2_{K_2}&=&     2(M_{\Q_2} \cos \phi_2 +M_{\Q_1} \cos \phi_1) g_* f\\
m_{\eta}^2&=& \frac{4}{3}(M_{\Q_2} \cos \phi_2+2 M_{\Q_1} \cos \phi_1) g_* f \ . \nonumber
\end{eqnarray}
the first solution is the global minimum at $\theta=0$ while the second is the minimum at $\theta=\pi$.
The two vacua are split for $\theta\ne \pi$ so that the higher minimum becomes a saddle point of the potential approaching $\theta\to0$.

\subsection{Examples of coset potentials}
We now turn to some explicit theories with $\SU(N_c)$ gauge group and lowest number of dark quarks, $N_F=2$ and $3$ and $\theta=0$.
We adopt the standard parametrisation  $\mathscr{U} = \exp (i \pi^a \lambda^a / f)$ with
${\rm Tr} \, \lambda^a \lambda^b = 2 \delta^{ab}$ such that  the `dark-pion' Goldstone boson $\pi^a$ have canonical normalization at the origin.

For $N_F=2$ the coset group is $\SU(2)$  with topology $S^3$, a sphere in 4 dimensions.
For $N_F=3$ the $\SU(3)$ coset has topology $S^3\times S^5$.\footnote{This can be seen defining 9 generators $\lambda_{ij}$ in terms of the $2\times 2$ Pauli matrices $\sigma_i$ of $\SU(2)$:
$\lambda_{ij}$ equals to $\sigma_i$ with extra zeroes in the $j$-th position.
%\beq (\lambda_{ij})_{IJ} = \left\{\begin{array}{ll}
%(\sigma_i)_{IJ} & \hbox{if $I,J\neq j$}\\
%0 & \hbox{otherwise}
%\end{array}\right.\eeq
Among the 9 generators of $\SU(2)^3$, one is redundant, merging $\SU(2)^2$ in a $S^5$.
The usual Gell-Mann basis is
$\lambda^1 = \lambda_{13}$,  $\lambda^2 = \lambda_{23}$, $\lambda^3 = \lambda_{33}$, 
$\lambda^4 = \lambda_{12}$,  $\lambda^5 = \lambda_{22}$, 
$\lambda^6 = \lambda_{11}$,  $\lambda^7 = \lambda_{21}$,
$\lambda^8 = (\lambda_{32}+\lambda_{31})/\sqrt{3}$.}
Its centers  $\mathscr{U}_n= e^{2\pi i N/3} \One$ for $n=\{0,1,2\}$ can be reached acting as 
$\exp(2\pi i n \lambda^8/\sqrt{3})$ on $\mathscr{U}=\One$.
Furthermore, extra points such as $\mathscr{U}=\diag(-1,-1,1)$ can be special for specific gauge and Yukawa interactions.

\subsubsection*{$N_F=2$, $\Q=1\oplus1$:}
The case of two dark-quarks charged under an ${\rm U}(1)$ gauge interaction is realised in QCD with the $u,d$ quarks
charged under electro-magnetism {(gauge generator $T = \text{diag}(2/3,-1/3)$)}.
The $\lambda^a$ reduce to the Pauli matrices $\sigma^a$
and the three dark-pions form a neutral $\pi^0$ and a charged $\pi^\pm$.
The coset matrix $\mathscr{U} = \exp (i \pi^a \sigma^a / f) = \One \cos\sfrac{\Pi}{f} + i  \sigma^a(\pi^a / \Pi) \sin\sfrac{\Pi}{f}$ 
can be computed analytically, in terms of
$\Pi^2 = \sum_a (\pi^a)^2 = (\pi^{0})^2+2\pi^+\pi^-$. 
The two elements of the center $\mathscr{U}_n = (-1)^n\One$ correspond to
$\Pi = 0$ and $\pi f$ and they are connected along the $\pi^3$ direction as $\mathscr{U} = \exp (i \pi^3 \sigma^3 / f)$,
with $\pi^3$ ranging between $0$ and $\pi f$.

The resulting potential is well known
%(see e.g.~\cite{1005.4269})
\begin{eqnsystem}{sys:VQCD}
V_{\rm mass} &=& -2 g_* f^3 (M_{\Q_1}+M_{\Q_2}) \cos\frac{\Pi}{f},  \label{eq:VmassQCD}\\
V_{\rm gauge} &=& \frac{3  g_*^2 f^2 }{(4\pi)^2} M_\gamma^2\qquad\hbox{with}\qquad
M_\gamma^2= e^2 f^2 \frac{\pi^+\pi^-}{\Pi^2}  \sin^2\frac{\Pi}{f}.
\end{eqnsystem}
%\LDL{[Se prendo il contributo in eq.~(16c) con $T^a = \text{diag}(2/3,-1/3)$ si ottiene
%\beq 
%V_{\rm gauge} = \frac {3  g_*^2 f^4 } {2(4\pi)^2} e^2 \frac{2 \pi^+ \pi^-}{\Pi^2} \sin^2\frac{\Pi}{f}
%\eeq
%forse si potrebbe scrivere questa forma consistente con eq.~(16c)]}
In QCD $V_{\rm mass}$ dominates over $V_{\rm gauge}$, such that the only minimum is at $\Pi=0$.
We consider a more general range of parameters, realised as
dark color with singlet $\Q$ possibly charged under hypercharge U(1)$_Y$.
$V_{\rm gauge}$ vanishes at $\Pi=0$ and $\Pi = \pi f$;
the two minima are separated by a barrier along the $\pi^\pm$ direction,
and are smoothly connected along the $\pi^0$ direction (analogously to the left panel of Fig.~\ref{fig:Palla}).

The potential $V_{\rm Yukawa}$ generated by possible Yukawa couplings of $\Q$ to scalars is model-dependent.
It can generate barriers, and it is flat along $\pi^0$ in models where its shift symmetry
corresponds to an U(1) accidental symmetry of the Yukawa interactions.

%\footnote{A similar expression holds in QCD,
%with $h$ replaced by $\pi^{\pm,0}$ (see e.g.~\cite{1005.4269,1403.3116}).  The term $V_1$ is generated by quark masses 
%$m_q$
%as $V_1/f^4 \sim m_q/f$
%and gives a common pion mass.
%QED generates $V_2/f^4 \sim \alpha_{\rm em} $
%and gives the smaller splitting $m^2_{\pi^\pm}-m^2_{\pi^0}$.
%The pion potential has only one minimum. \LDL{[Da rivedere/connettere con la discussione nella sez.~successiva]}}

\subsubsection*{$N_F=2$,  $\Q=2$:}

Alternatively, the fermions $\Q$ can form a doublet under $\SU(2)_L$ with hypercharge $Y$.
The dark pions form a $\SU(2)_L$ triplet with zero hypercharge. Thereby hypercharge does not  contribute to the gauge potential
\beq 
V_{\rm gauge} = \frac{3  g_*^2 f^4 }{(4\pi)^2} g_2^2 \sin^2\frac{\Pi}{f}
%\qquad\hbox{with}\qquad 
%M_V^2= g^2 {f^2} \sin^2\frac{\Pi}{f}
\eeq
which contains two inequivalent degenerate minima separated by potential barriers.
The mass potential is obtained from eq.\eq{VmassQCD} setting degenerate $\Q$ masses,
$V_{\rm mass} = -4 g_* f^3 M_{\Q} \cos\sfrac{\Pi}{f}$.
It splits the two minima, possibly removing one of them if
$V_{\rm mass}$ dominates over $V_{\rm gauge}$.

\subsubsection*{$N_F=3$,  $\Q=3$:}\label{Q=3}

Assuming that $\Q$ is a triplet of $\SU(2)_L$ leads to a dark-matter model~\cite{1503.08749}.
The dark-pions $\pi^a$ have zero hypercharge and decompose as $3\oplus 5  = \vec\pi_3\oplus\vec\pi_5$ under $\SU(2)_L$,
with $\vec\pi_5=\{\pi^1,\pi^3,\pi^4,\pi^6,\pi^8\}$ and
$\vec\pi_3=\{\pi^2,\pi^5,\pi^7\}$ containing a stable dark-matter candidate 
(dark-baryons provide an extra dark-matter candidate,
if $\Q$ has zero hypercharge).
The $\SU(2)_L$ generators are
$ T^b_3 =  \{ \lambda^{2}, \lambda^{5}, \lambda^{7} \}$.
%while the fiveplet with $\vec T_5 = \tfrac{1}{2} \( \lambda^{1}, \lambda^{3}, \lambda^{4}, \lambda^{6}, \lambda^{8} \)$.}
% questo non ha senso
%$\vec\pi_3$ \LDL{\sout{is} 
Neither $\mathscr{U}$ nor the gauge potential $V_{\rm gauge}(\vec\pi_3, \vec\pi_5) $ can be written in an useful closed form.
For $\vec\pi_5=0$ it equals 
\be
V_{\rm gauge} (\vec\pi_3,\vec\pi_5=0) =  -\frac {6  g_*^2 f^4 } {(4\pi)^2} g^2_2
\cos\frac{\sqrt{\vec\pi^2_3}}{f}
\ee
and is minimal at the origin $\vec\pi_3=0$.  
Turning on only $\vec\pi_5$ the potential
does not depend only on $\vec\pi_5^2$, and has different periodicity along its $\pi^8$ component.
The potential along $\pi^8$, with all other components vanishing 
%\LDL{
%\sout{
%%\begin{equation}
%$V_{\rm gauge} =\xxx{?}   -\frac13 \[1+2\cos\frac{2\sqrt{3}\pi_8}{f}\]$
%%\end{equation}
%}
\be
V_{\rm gauge}  = -\frac{12 g_*^2 f^4}{(4\pi)^2} g_2^2 
\cos\frac{\sqrt{3}\pi^8}{f} 
\ee
has three degenerate minima at $ \pi^8_n = 2 \pi n/ \sqrt{3}$ with $n=\{-1,0,1\}$
in correspondence of the centers 
$\mathscr{U}_n = \exp (i \pi^8_n \lambda^8 / f)$, 
%\DT{Non capisco, il centro \`e per valori discreti}
%$\mathscr{U}_n=e^{2\pi i n /3}\diag(1,1,1)$}, 
separated by potential barriers.
A numerical study shows that these are the only local minima. 
The potential due to constituent masses
\begin{equation}
V_{\rm mass}= - 2 g_* f^3 M_{\Q}
\[2\cos\frac{\pi^8}{\sqrt{3}f} +  \cos\frac{2\pi^8}{\sqrt{3}f} \]
\end{equation} 
makes the origin deeper than the 
 the other two centers for
$M_{\Q}>0$.

% $V_{\rm mass}$ is minimal at $N=0$ and maximal at $N=\{1,2\}$.

%and shifts the potential at the centers, providing an extra barrier only between the minima
%with $N=2,3$ which remain degenerate
%
%For $M_3>0$ $\mathscr{U}=\One$ becomes the only minimum;
%for $M_3<0$ \xxx{makes sense?} the two other centers become degenerate minima.

%\begin{equation}
%\Delta \sim m g_* f^3 
%\end{equation}
%When the masses dominate the metastable minima disappear.

\subsubsection*{$N_F=3$, $\Q=2\oplus 1$}
\label{Q=2+1}
The dark-pions $\pi^a$ decompose as $3\oplus 2\oplus \bar 2 \oplus 1$ under $\SU(2)_L$.
The singlet $\pi^8$
%\LDL{[per simmetria con la discussione precedente lo chiamerei $\pi^8$]} 
has zero hypercharge irrespectively of the unspecified hypercharges of the two $\Q$ components.
%The $3$ parameterize $S^3$, and $2\oplus \bar 2 \oplus 1$ parameterizes $S^5$?
Local minima of $V_{\rm gauge}$
have the form $\mathscr{U}={\rm diag}(e^{i\alpha}, e^{i\alpha},e^{-2 i \alpha})$,
given that the $\SU(2)_L$ generators $T^b$ act on the first two components. %Some points can be reached turning on the Higgs $2\oplus \bar 2$.
All the minima are of the form  $\mathscr{U}=\exp(i \pi^8 \lambda^8/f)$. 
It is convenient to introduce  $\eta  \equiv  \pi^8/\sqrt{3}$ (non-canonically normalized at the origin) with periodicity $2\pi f$.
The antipodal point  $\mathscr{U}={\rm diag}(-1,-1,1)$ is obtained for 
$\eta/f=\pi$;
the elements of the center $\mathscr{U}_{n} = e^{2 \pi i n/3} \One$ are obtained for $\eta/f = 2 \pi n/3$.
The flatness of $V_{\rm gauge}$ along $\eta$ is lifted by the potential generated by constituent masses:
\be \label{eq:VmassLN}
 V_{\rm mass} = -4 g_* f^3 M_{\Q_2} \cos \frac{\eta}{f} - 2 g_* f^3 M_{\Q_1} \cos \frac{2 \eta}{f}
\ee
when setting all the other Goldstone bosons to zero. 
$V_{\rm mass}$ has a minimum at $\eta=0$ and at $\eta = \pi f$ for $2M_{\Q_1}>M_{\Q_2}>0$.
The minima at $\mathscr{U} = \diag(1,1,1)$ and $\mathscr{U} = \diag(-1,-1,1)$ are split by $M_{\Q_2}$ and a barrier between them in the full potential is created by  $M_{\Q_1}$. 
%The centers 
%$\mathscr{U}_{1,2} = e^{2 \pi i N/3} \One$ with $N=1,2$ tend to disappear when $V_{\rm mass}$ dominates.
%Luca, Vmass ha il  minimo esattamente \pi f; ottieni un minimo che si sposta se combini Vmass con altri contributi, ad esempio il gauge.

Constituent masses and Yukawa interactions that respect the $\mathbb{Z}_3$ symmetry between the centers 
can potentially realise the tri-phase scenario of~\cite{1802.10093}.

\end{document}